\documentclass[aps,prd,twocolumn,floatfix,preprintnumbers,nofootinbib,superscriptaddress]{revtex4}
\usepackage{amsmath,amsthm,amssymb,amsfonts}
\usepackage{graphicx}
\usepackage{textcomp}
\usepackage{bm}
\usepackage{color}
\usepackage{multirow}
\usepackage{dutchcal}
\usepackage{makecell}

\usepackage[outdir=./]{epstopdf}

\begin{document}

\title{Theoretical study on low-lying hidden-bottom and double-bottom tetraquark states}
	
	\author{Lin-Juan Jiang}
	\affiliation{School of Physical Science and Technology, Southwest
		University, Chongqing 400715, China}
	
	\author{Chun-Sheng An}~\email{ancs@swu.edu.cn}
	\affiliation{School of Physical Science and Technology, Southwest
		University, Chongqing 400715, China}
	
	\author{Cheng-Rong Deng}~\email{crdeng@swu.edu.cn}
	\affiliation{School of Physical Science and Technology, Southwest
		University, Chongqing 400715, China}
	
	\author{Gang Li}~\email{gli@qfnu.edu.cn}
	\affiliation{College of Physics and Engineering, Qufu Normal
		University, Qufu 273165, China}
	
	\author{Ju-Jun Xie}~\email{xiejujun@impcas.ac.cn}
	\affiliation{Institute of Modern Physics, Chinese Academy of Sciences, Lanzhou 730000, China} 
	\affiliation{School of Nuclear Science and Technology, University of Chinese Academy of Sciences, Beijing 101408, China} \affiliation{Southern Center for Nuclear-Science Theory (SCNT), Institute of Modern Physics, Chinese Academy of Sciences, Huizhou 516000, China}
	\date{\today}
	
	\begin{abstract}
			
We perform a theoretical  study on the spectrum of the low-lying hidden-bottom ($q\bar{q}b\bar{b}$ with $q=$ $u$, $d$ and $s$) and double-bottom ($q\bar{q}bb$) tetraquark states within a nonrelativistic quark model, in which the instanton-induced interaction is taken as the residual spin-dependent hyperfine interaction between quarks. All the model parameters are fixed by fitting the spectrum of the ground hadron states. The numerical results indicate that masses of several $X_{q\bar{q}}$, $Z_{b}$, and $Z_{bs}$ tetraquark states are below and near thresholds of corresponding meson-meson channels, thus these states may form components of exotic meson states with reasonable probabilities. Especially, in present model, masses of several obtained states with quantum number $I^G(J^P)=1^+(1^{+})$ are close to $Z_b^\pm(10610)$, so one may expect these states to be non-negligible components of the experimentally observed $Z_{b}$ states. Concerning to the double-bottom tetraquark states, the present results are in general consistent with other previous works. Two possible stable $T_{bb}$ states with quark content $bb\bar{n}\bar{n}$ ($n=u$ or $d$ quark) lying at energies $10558$~MeV and $10650$~MeV are found, and one possible stable $bb\bar{n}\bar{s}$ state is found, whose energy is $\sim10687$~MeV.

	\end{abstract}
	
	\maketitle

\section{Introduction}
\label{Introduction}

In the last two decades, experimental and theoretical investigations on the exotic meson states are one of the most interesting topics in hadronic physics since the observation of $X(3872)$~\cite{Belle:2003nnu}. Concerning to the meson states with bottom quarks, in 2011, Belle collaboration firstly reported their results on measurement of the processes $e^+e^-\to \Upsilon(nS)\pi^+\pi^-$ and $e^+e^-\to h_b(mP)\pi^+\pi^-$ in ``the 9th Conference on Flavor Physics and CP Violation" (FPCP 2011), it's shown that in the invariant mass of $\pi^\pm\Upsilon(nS)(n=1,2,3)$ and $\pi^\pm h_b(mP)(m=1,2)$, two charged bottomonium-like states $Z_b^\pm(10610)$ and $Z_b^\pm(10650)$ should exist~\cite{Adachi:2011mks}. Obviously, these two states cannot be explained as traditional meson states. Later, the explicit measured masses and decay widths averaged over the five different final states, $M_{Z_b(10610)}=10607.2\pm2.0$~MeV, $M_{Z_b(10650)}=10652.2\pm1.5$~MeV, $\Gamma_{Z_b(10610)}= 18.4\pm2.4$~MeV, $\Gamma_{Z_b(10650)} = 11.5\pm2.2$~MeV, are published in Ref.~\cite{Belle:2011aa}, where the angular distribution analysis indicates that the quantum numbers of both $Z_b^\pm(10610)$ and $Z_b^\pm(10650)$ should be $I^G(J^P)=1^+(1^+)$. Accordingly, on the theoretical side, intensive investigations on the structure of $Z_b$ states have been done, in which the observed $Z_b^\pm(10610)$ and $Z_b^\pm(10650)$ have been explained as tetraquark states ~\cite{Ali:2011ug,Cui:2011fj,Agaev:2017lmc,Wang:2013zra,Wang:2019mxn}, molecular states~\cite{Mehen:2011yh, Ohkoda:2011vj,Dong:2012hc,Ke:2012gm,Bondar:2011ev,Voloshin:2011qa, Nieves:2011zz,Cleven:2011gp,Zhang:2011jja,Wang:2013daa,Wang:2014gwa}, threshold cusps~\cite{Bugg:2011jr}, or re-scattering effects~\cite{Chen:2011zv,Li:2012as}, etc.
	
On the other hand, a doubly heavy baryon $\Xi^{++}_{cc}$ was observed by the LHCb Collaboration~\cite{LHCb:2017iph}, which could provide an excellent opportunity to examine the interactions between two heavy quarks and search for more doubly heavy quark states. Meanwhile, based on the mass of $\Xi^{++}_{cc}$, the mass spectra of doubly bottom tetraquark states $T_{bb}$ were studied subsequently, which indicated that there should exist at least one stable flavored exotic tetraquark $bb\bar{u}\bar{d}$~\cite{Karliner:2017qjm,Eichten:2017ffp}. Actually, various approaches have been applied to the double-bottom tetraquark states, including the chromomagnetic interaction (CMI) model~\cite{Cheng:2020wxa,Luo:2017eub}, quark-level models ~\cite{Semay:1994ht,Pepin:1996id,Brink:1998as,Ebert:2007rn,Zhang:2007mu,Yang:2009zzp,Feng:2013kea,Karliner:2017qjm,Eichten:2017ffp,Park:2018wjk,Deng:2018kly,Bedolla:2019zwg,Hernandez:2019eox,Yang:2019itm,Yu:2019sxx,Lu:2020rog,Tan:2020ldi,Yang:2020fou,Weng:2021hje,Song:2023izj,Zhang:2021yul}, QCD sum rule method~\cite{Wang:2010uf,Du:2012wp,Wang:2017uld,Tang:2019nwv,Agaev:2019lwh,Navarra:2007yw,Gao:2020bvl}, lattice QCD simulation ~\cite{Brown:2012tm,Bicudo:2015vta,Francis:2016hui,Francis:2018jyb,Junnarkar:2018twb,Leskovec:2019ioa,Hudspith:2020tdf,Braaten:2020nwp}, etc. 

More investigations on the hidden-bottom and double-bottom tetraquarks from the different theoretical perspectives must benefit the future experimental exploration of the tetraquark states. The constituent quark model (CQM) is one of the most successful phenomenological methods to study the hadron structure. In present work, spectrum of the low-lying hidden- and double-bottom tetraquark states are investigated in the quark potential model, in which the instanton-induced interaction is taken as the residual spin-dependent hyperfine interaction between quarks. In fact, the instanton-induced interaction has already been successfully applied to the traditional hadron spectroscopy~\cite{Shuryak:1988bf,Blask:1990ez,Brau:1998sxe,Semay:2001th}, tetraquark states~\cite{Beinker:1995qe,tHooft:2008rus,Wang:2022clw,Zhang:2022qtp}, as well the transition couplings between traditional and exotic baryon states~\cite{An:2013zoa}.
	
The present manuscript is organized as follows. In Sec.~\ref{Framework}, we present the theoretical formalism, including effective Hamiltonian with instanton-induced interactions between quarks, wave function of the studied tetraquark states, and the model parameters. Numerical results and discussions for the spectrum of the hidden-bottom and double-bottom tetraquarks are shown in Sec.~\ref{Results and Discussions}. Finally, Sec.~\ref{Summary} contains a brief conclusion.

\section{Theoretical Framework}
\label{Framework}

\subsection{Effective Hamiltonian}

In present work, the nonrelativistic quark potential model is used for calculations on the spectrum of the studied tetraquark spectrum. In the quark model, the effective Hamiltonian can be expressed as:
	\begin{equation}
		H_{eff.}=\sum_{i=1}^{4}\left(m_i+T_i\right)-T_{C.M.}+V_{Conf.}+V_{Ins.}\,,\label{H}
	\end{equation}
where $m_i$ and $T_i$ denote the constituent mass and kinetic energy of $i$-th quark, $T_{C.M.}$ is the center of mass kinetic energy. And $V_{Conf.}$ is the quark confinement potential, which reads~\cite{Wang:2022clw,Zhang:2022qtp,Eichten:1974af}:
	\begin{equation}
		V_{Conf.}=\sum_{i<j}-\frac{3}{16}\,\left(\vec{\lambda}^c_i\cdot\vec{\lambda}^c_j\right)\,\left(b\,r_{ij}-\frac{4}{3}\frac{\alpha_{ij}}{r_{ij}}+C_0\right)\,,
	\end{equation}
where $\vec{\lambda}^c{_i(_j)}$ is Gell-Mann matrix in $SU(3)$ color space acting on the $i(j)$-th quark, $b$,  $\alpha_{ij}$ and $C_0$  are strength of quark confinement, QCD effective coupling constant between two quarks and zero point energy, respectively.
	
For the residual spin-dependent interaction, here the phenomenological extended version of 't Hooft's Instanton-induced interaction is employed~\cite{Migura:2006ep}:
	\begin{equation}
		V_{Ins.}=V^{qq}_{Ins.}+V^{q\bar{q}}_{Ins.}\,,
	\end{equation}
with
	\begin{align}
		V^{qq}_{Ins.}=&\sum_{i<j}-\hat{g}^{qq}_{ij}\left(P_{ij}^{S=1}P_{ij}^{C,\bf{6}}+2P_{ij}^{S=0}P_{ij}^{C,\bf{\bar{3}}}\right)\delta^3\left(\vec{r}_{ij}\right)\,,\\
		V^{q\bar{q}}_{Ins.}=&\sum_{i<j}\hat{g}^{q\bar{q}}_{ij}\left[\frac{3}{2}P_{ij}^{S=1}P_{ij}^{C,\bf{8}}+P_{ij}^{S=0}\left(\frac{1}{2}P_{ij}^{C,\bf{8}}\right.\right.\notag\\
		&\left.\left.+8P_{ij}^{C,\bf{1}}\right)\right]\delta^3\left(\vec{r}_{ij}\right)\,,
	\end{align}
where $V^{qq}_{Ins.}$ stands for the interactions between quark-quark pair or antiquark-antiquark pair, while $V^{q\bar{q}}_{Ins.}$ stands for those between the quark-antiquark pair. $\hat{g}^{qq}_{ij}$ and $\hat{g}^{q\bar{q}}_{ij}$ are flavor-dependent coupling strength operators, the explicit matrix elements of these operators in the light quark, strange quark and charm quark sectors have been shown in Ref.~\cite{Wang:2022clw}, concerning to the bottom sector, the new involved matrix elements are very similar. One may notice that the matrix elements of $\hat{g}^{q\bar{q}}_{ij}$ between two $q\bar{q}$ pairs with different flavors are nonzero~\cite{Wang:2022clw}, thus they will lead to mixing between different tetraquark states, for instance, the $b\bar{b}u\bar{u}$ and $b\bar{b}s\bar{s}$ states studied in present work, once the transition matrix elements of $\hat{g}^{q\bar{q}}_{ij}$ are obtained, mixing coefficients between $b\bar{b}u\bar{u}$ and $b\bar{b}s\bar{s}$ configurations can be directly calculated by diagonalization of the energy matrix. $P_{ij}^{S=0}$, $P_{ij}^{S=1}$ are spin projector operators onto spin singlet and spin triplet states, respectively, finally, $P_{ij}^{C,\bf{\bar{3}}}$, $P_{ij}^{C,\bf{6}}$, $P_{ij}^{C,\bf{1}}$ and $P_{ij}^{C,\bf{8}}$ are color projector operators onto color anti-triplet $\bf{\bar{3}}_c$, color sextet $\bf{6}_c$, color singlet $\bf{1}_c$ and color octet $\bf{8}_c$, respectively.

As we know, the pure contact interaction term with $\delta^3\left(\vec{r}_{ij}\right)$ should lead to an unbound Hamiltonian, therefore, here the delta function in the instanton-induced interaction is regularized using the approach proposed in Refs.~\cite{Godfrey:1985xj,Vijande:2004he,Beinker:1995qe},
	\begin{equation}
		\delta^3\left(\vec{r}_{ij}\right)\,\rightarrow\,\left(\frac{\sigma}{\sqrt{\pi}}\right)^3\mathrm{exp}\left(-\sigma^2\,r_{ij}^2\right)\,,
	\end{equation}
where $\sigma$ is a regularization parameter, which should depend on the finite size of constituent quarks.

\subsection{Wave Functions}

To calculate the spectroscopy of hidden- and double-bottom tetraquark states system, first we construct wave functions of the tetraquark configurations in the space of spin$\otimes$color$\otimes$flavor. In general, there are eight different kinds of symmetric configurations for $qq\bar{q}\bar{q}$:
	\begin{flalign}
		\hspace{1cm}|1\rangle&=\{qq\}_{{\bf{6}}_c}\{\bar{q}\bar{q}\}_{\bar{\bf{6}}_c}\,, & |2\rangle&=[qq]_{\bar{\bf{3}}_c}[\bar{q}\bar{q}]_{{\bf{3}}_c}\,,\hspace{1cm}\notag\\
		|3\rangle&=\{qq\}^{*}_{\bar{\bf{3}}_c}[\bar{q}\bar{q}]_{{\bf{3}}_c}\,,& |4\rangle&=[qq]_{\bar{\bf{3}}_c}\{\bar{q}\bar{q}\}^{*}_{{\bf{3}}_c}\,,\notag\\
		|5\rangle&=[qq]^{*}_{{\bf{6}}_c}\{\bar{q}\bar{q}\}_{\bar{\bf{6}}_c}\,,& |6\rangle&=\{qq\}_{{\bf{6}}_c}[\bar{q}\bar{q}]^{*}_{\bar{\bf{6}}_c}\,,\notag\\
		|7\rangle&=\{qq\}^{*}_{\bar{\bf{3}}_c}\{\bar{q}\bar{q}\}^{*}_{{\bf{3}}_c}\,,& |8\rangle&=[qq]^{*}_{{\bf{6}}_c}[\bar{q}\bar{q}]^{*}_{\bar{\bf{6}}_c}\,,
	\end{flalign}
where the superscript $\ast$ means diquark or antidiquark forms a spin triplet, while the configurations without superscript are spin singlet. The blackened number in the subscript denotes color wave function of two-quark (-antiquark) subsystem. Finally, $\{~\}$ and $[~]$ denote the symmetric and anti-symmetric flavor wave functions of the two quarks (antiquarks) subsystems, respectively.

Explicit spin, color, flavor, coordinate space wave functions of the hidden- and double-bottom tetraquark systems are presented in the following subsections, respectively.

\subsubsection{Spin wave function}

For the spin part, six spin wave functions can be obtained from the wave function of the two quarks $qq$ and $\bar{q}\bar{q})$:
	\begin{flalign}
		\chi_0^{00}&=\vert(q_1q_2)_0(\bar{q_3}\bar{q_4})_0\rangle_0, ~~~
		\chi_0^{11}=\vert(q_1q_2)_1(\bar{q_3}\bar{q_4})_1\rangle_0, \notag\\
		\chi_0^{01}&=\vert(q_1q_2)_0(\bar{q_3}\bar{q_4})_1\rangle_1, ~~~
		\chi_0^{10}=\vert(q_1q_2)_1(\bar{q_3}\bar{q_4})_0\rangle_1, \notag\\
		\chi_0^{11}&=\vert(q_1q_2)_1(\bar{q_3}\bar{q_4})_1\rangle_1, ~~~
		\chi_0^{11}=\vert(q_1q_2)_1(\bar{q_3}\bar{q_4})_1\rangle_2,
	\end{flalign}
where the subscripts denote the spin quantum numbers of the diquark, the antidiquark, and the tetraquark state. Explicit expressions of the above spin wave functions are as follows:
    \begin{eqnarray} \chi_0^{00}&=&\frac{1}{2}(\uparrow\downarrow\uparrow\downarrow-\uparrow\downarrow\downarrow\uparrow
    -\downarrow\uparrow\uparrow\downarrow+\downarrow\uparrow\downarrow\uparrow)\, \nonumber \\
    \chi_0^{11} &=& \sqrt{\frac{1}{12}}(2\uparrow\uparrow\downarrow\downarrow-\uparrow\downarrow\uparrow\downarrow
    -\uparrow\downarrow\downarrow\uparrow-\downarrow\uparrow\uparrow\downarrow-\downarrow\uparrow\downarrow\uparrow\nonumber\\
    &&+2\downarrow\downarrow\uparrow\uparrow)\,,\nonumber\\
	\chi_1^{01}&=&\sqrt{\frac{1}{2}}(\uparrow\downarrow\uparrow\uparrow-\downarrow\uparrow\uparrow\uparrow)\,,\nonumber\\
	\chi_1^{10}&=&\sqrt{\frac{1}{2}}(\uparrow\uparrow\uparrow\downarrow-\uparrow\uparrow\downarrow\uparrow)\,,\nonumber\\ \chi_1^{11}&=&\frac{1}{2}(\uparrow\uparrow\uparrow\downarrow+\uparrow\uparrow\downarrow\uparrow-\uparrow\downarrow\uparrow\uparrow
-\downarrow\uparrow\uparrow\uparrow)\,,\nonumber\\
	\chi_2^{11}&=&\uparrow\uparrow\uparrow\uparrow\,.
\end{eqnarray}
With the above spin wave functions, one can work out the matrix elements of corresponding spin operators listed in Table~\ref{Spin-matrix-elements}.
\begin{table}[htbp]
	\caption{Spin matrix elements.} \label{Spin-matrix-elements}
	\renewcommand
	\tabcolsep{0.02cm}
	\renewcommand{\arraystretch}{1.8}
	\begin{tabular}{ccccccc}
		\hline\hline
		          &$\langle\sigma_1\cdot\sigma_2\rangle$&$\langle\sigma_1\cdot\sigma_3\rangle$&$\langle\sigma_1\cdot\sigma_4\rangle$&$\langle\sigma_2\cdot\sigma_3\rangle$&$\langle\sigma_2\cdot\sigma_4\rangle$&$\langle\sigma_3\cdot\sigma_4\rangle$\\\hline
		$\langle\chi_0^{00}\vert\hat{O}\vert\chi_0^{00}\rangle$&$-3$&$0$&$0$&$0$&$0$&$-3$\\
		$\langle\chi_0^{11}\vert\hat{O}\vert\chi_0^{11}\rangle$&$1$&$-2$&$-2$&$-2$&$-2$&$1$\\
		$\langle\chi_0^{00}\vert\hat{O}\vert\chi_0^{11}\rangle$&$0$&$-\sqrt{3}$&$\sqrt{3}$&$\sqrt{3}$&$-\sqrt{3}$&$1$\\
		$\langle\chi_1^{01}\vert\hat{O}\vert\chi_1^{01}\rangle$&$-3$&$0$&$0$&$0$&$0$&$1$\\
		$\langle\chi_1^{10}\vert\hat{O}\vert\chi_1^{10}\rangle$&$1$&$0$&$0$&$0$&$0$&$-3$\\
		$\langle\chi_1^{11}\vert\hat{O}\vert\chi_1^{11}\rangle$&$1$&$-1$&$-1$&$-1$&$-1$&$1$\\
		$\langle\chi_1^{01}\vert\hat{O}\vert\chi_1^{11}\rangle$&$0$&$1$&$-1$&$-1$&$1$&$0$\\
		$\langle\chi_0^{01}\vert\hat{O}\vert\chi_1^{11}\rangle$&$0$&$-\sqrt{2}$&$-\sqrt{2}$&$\sqrt{2}$&$\sqrt{2}$&$1$\\
		$\langle\chi_0^{10}\vert\hat{O}\vert\chi_1^{11}\rangle$&$0$&$\sqrt{2}$&$-\sqrt{2}$&$\sqrt{2}$&$-\sqrt{2}$&$1$\\
		$\langle\chi_2^{11}\vert\hat{O}\vert\chi_2^{11}\rangle$&$1$&$1$&$1$&$1$&$1$&$1$\\
			\hline\hline
	\end{tabular}
\end{table}

\subsubsection{Color wave function}

The color wave function can be obtained by using the SU(3) group theory, the direct product of the diquark–antidiquark components reads $3_c\otimes3_c\otimes\bar{3_c}\otimes\bar{3_c}=(6_c\oplus\bar{3_c})\otimes(\bar{6_c}\oplus3_c)$. Accordingly, to form the required color singlet, one can get the following two wave functions:
	\begin{flalign}
		\zeta_1=6_c\otimes\bar{6_c}=\vert(q_1q_2)^6(\bar{q_3}\bar{q_4})^{\bar{6}}\rangle&,\notag\\
		\zeta_2=\bar{3_c}\otimes3_c=\vert(q_1q_2)^{\bar{3}}(\bar{q_3}\bar{q_4})^3\rangle&.
	\end{flalign}
Explicit forms of the above two color wave functions are as follows:
    	\begin{eqnarray}	\zeta_1&=&\frac{1}{2\sqrt{6}}(2rr\bar{r}\bar{r}+2gg\bar{g}\bar{g}+2bb\bar{b}\bar{b}+rg\bar{r}\bar{g}+rg\bar{g}\bar{r}\nonumber\\
    &&+gr\bar{g}\bar{r}+gr\bar{r}\bar{g}+rb\bar{r}\bar{b}+rb\bar{b}\bar{r}+br\bar{b}\bar{r}\nonumber\\
    &&+br\bar{r}\bar{b}+gb\bar{g}\bar{b}+gb\bar{b}\bar{g}+bg\bar{b}\bar{g}+bg\bar{g}\bar{b})\,,\nonumber\\
\zeta_1&=&\frac{1}{2\sqrt{3}}(rg\bar{r}\bar{g}-rg\bar{g}\bar{r}+gr\bar{g}\bar{r}-gr\bar{r}\bar{g}+rb\bar{r}\bar{b}-rb\bar{b}\bar{r}\nonumber\\
&&+br\bar{b}\bar{r}-br\bar{r}\bar{b}+gb\bar{g}\bar{b}-gb\bar{b}\bar{g}+bg\bar{b}\bar{g}-bg\bar{g}\bar{b})\,.
    	\end{eqnarray}	
With these color wave functions, one can work out the matrix elements of the corresponding color operators listed in TABLE~\ref{Color-matrix-elements}.
\begin{table}[htbp]
	\caption{Color matrix elements.} \label{Color-matrix-elements}
	\renewcommand
	\tabcolsep{0.02cm}
	\renewcommand{\arraystretch}{1.8}
	\begin{tabular}{ccccccc}
		\hline\hline
		        &$\langle\lambda_1\cdot\lambda_2\rangle$&$\langle\lambda_1\cdot\lambda_3\rangle$&$\langle\lambda_1\cdot\lambda_4\rangle$&$\langle\lambda_2\cdot\lambda_3\rangle$&$\langle\lambda_2\cdot\lambda_4\rangle$&$\langle\lambda_3\cdot\lambda_4\rangle$\\\hline
		        $\langle\zeta_1\vert\hat{O}\vert\zeta_1\rangle$&$\frac{4}{3}$&$-\frac{10}{3}$&$-\frac{10}{3}$&$-\frac{10}{3}$&$-\frac{10}{3}$&$\frac{4}{3}$\\
		        $\langle\zeta_2\vert\hat{O}\vert\zeta_2\rangle$&$-\frac{8}{3}$&$-\frac{4}{3}$&$-\frac{4}{3}$&$-\frac{4}{3}$&$-\frac{4}{3}$&$-\frac{8}{3}$\\
		        $\langle\zeta_1\vert\hat{O}\vert\zeta_2\rangle$&$0$&$-2\sqrt{2}$&$2\sqrt{2}$&$2\sqrt{2}$&$-2\sqrt{2}$&$0$\\
		\hline\hline
	\end{tabular}
\end{table}

\subsubsection{Flavor wave function}

For the hidden-bottom Systems, there are nine flavor configurations, those are
$b\bar{b}u\bar{u}$, $b\bar{b}u\bar{d}$, $b\bar{b}u\bar{s}$, $b\bar{b}d\bar{u}$, $b\bar{b}d\bar{d}$, $b\bar{b}d\bar{s}$, $b\bar{b}s\bar{u}$, $b\bar{b}s\bar{d}$, $b\bar{b}s\bar{s}$. Note that $b\bar{b}d\bar{u}$, $b\bar{b}s\bar{u}$, $b\bar{b}s\bar{d}$ are the antiparticles of $b\bar{b}u\bar{d}$, $b\bar{b}u\bar{s}$, $b\bar{b}d\bar{s}$, respectively. Therefore we only need to consider six configurations, i.e., $b\bar{b}u\bar{u}$, $b\bar{b}u\bar{d}$, $b\bar{b}u\bar{s}$, $b\bar{b}d\bar{d}$, $b\bar{b}d\bar{s}$, $b\bar{b}s\bar{s}$.

The six flavor configurations can be decomposed according to  isospin $I$ and strangeness $S$ as below:
\begin{itemize}
	\item $I=0,\,S=0.$
	\begin{align}
		X_{n\bar{n}}&=\frac{1}{\sqrt{2}}\left(b\bar{b}u\bar{u}+b\bar{b}d\bar{d}\right),\label{X1f}\\
		X_{s\bar{s}}&=b\bar{b}s\bar{s}.\label{X2f}
	\end{align}
	\item $I=\frac{1}{2},\,S=\pm1.$
	\begin{align}
		Z_{bs}^{+}=b\bar{b}u\bar{s},~~\bar{Z}_{bs}^{0}=-b\bar{b}s\bar{d},\label{fZbs+}\\
		Z_{bs}^{0}=b\bar{b}d\bar{s},~~Z_{bs}^{-}=b\bar{b}s\bar{u}.\label{fZbs0}
	\end{align}
	\item $I=1,\,S=0.$
	\begin{align}
		Z_{b}^{+}&=-b\bar{b}u\bar{d},\label{Zb+}\\
		Z_{b}^{0}&=\frac{1}{\sqrt{2}}\left(b\bar{b}u\bar{u}-b\bar{b}d\bar{d}\right),\label{Zb0}\\
		Z_{b}^{-}&=b\bar{b}d\bar{u}.\label{Zb-}
	\end{align}
\end{itemize}
The $X_{q\bar{q}}$ and $Z_b^0$ states are charge neutral systems, thus one can also categorize them by the C-parity quantum numbers, as shown in Table~\ref{Configurations-of-hidden-bottom-Systems}.
\begin{table}[h]
	\caption{Configurations of hidden-bottom systems.}\label{Configurations-of-hidden-bottom-Systems}
	\renewcommand
	\tabcolsep{0.3cm}
	\renewcommand{\arraystretch}{1.8}
	\begin{tabular}{cccc}
		\hline\hline
		\multicolumn{2}{c}{$Z_{bs}$}&\multicolumn{2}{c}{$X_{q\bar{q}}$ and $Z_b$} \\\hline
		
		$J^{P}$&Configuration&$J^{PC}$&Configuration\\\hline
		
		$0^+ $&  $|1\rangle$ &$0^{++} $&  $|1\rangle$ \\
		
		&  $|2\rangle$ &      &  $|2\rangle$ \\
		
		&  $|7\rangle$ &      &  $|7\rangle$ \\
		
		&  $|8\rangle$ &      &  $|8\rangle$ \\\hline
		
		$1^+$ &  $|3\rangle$ &$1^{++}$ &  $|3'\rangle=\frac{1}{\sqrt{2}}\left(|3\rangle+|4\rangle\right)$ \\
		
		&  $|4\rangle$ &      &  $|5'\rangle=\frac{1}{\sqrt{2}}\left(|5\rangle+|6\rangle\right)$ \\\cline{3-4}
		
		&  $|5\rangle$ &$1^{+-}$&  $|4'\rangle=\frac{1}{\sqrt{2}}\left(|3\rangle-|4\rangle\right)$ \\
		
		&  $|6\rangle$ &      &  $|6'\rangle=\frac{1}{\sqrt{2}}\left(|5\rangle-|6\rangle\right)$ \\
		
		&  $|7\rangle$ &      &  $|7\rangle$ \\
		
		&  $|8\rangle$ &      &  $|8\rangle$ \\\hline
		
		$2^+$ &  $|7\rangle$ &$2^{++}$&  $|7\rangle$ \\
		
		&  $|8\rangle$ &      &  $|8\rangle$ \\
		\hline\hline
	\end{tabular}
\end{table}

Similarly, for the double-bottom systems, there are also nine different kinds of flavor configurations, including the $bb\bar{u}\bar{u}$, $bb\bar{d}\bar{d}$, $bb\bar{s}\bar{s}$, $bb\bar{u}\bar{d}$/$bb\bar{d}\bar{u}$, $bb\bar{u}\bar{s}$/$bb\bar{s}\bar{u}$, $bb\bar{d}\bar{s}$/$bb\bar{s}\bar{d}$, which can be categorized based on isospin $I$ and strangeness $S$:
\begin{itemize}
	\item $I=0,S=0.$
	\begin{equation}
		\left(T_{bb}^{0,0}\right)^-=\frac{1}{\sqrt{2}}bb\left(\bar{u}\bar{d}-\bar{d}\bar{u}\right),\label{Tbb1}
	\end{equation}

	\item $I=0,S=2.$
	\begin{equation}
		\left(T_{bb}^{0,2}\right)^{0}=bb\bar{s}\bar{s},\label{Tbb2}
	\end{equation}

	\item $I=\frac{1}{2},S=1.$
	\begin{align}
		\left(T_{bb}^{\frac{1}{2},1}\right)^0&=-\frac{1}{\sqrt{2}}bb\left(\bar{d}\bar{s}\pm \bar{s}\bar{d}\right),\label{Tbb3}\\
		\left(T_{bb}^{\frac{1}{2},1}\right)^-&=\frac{1}{\sqrt{2}}bb\left(\bar{u}\bar{s}\pm \bar{s}\bar{u}\right),\label{Tbb4}
	\end{align}
	
	\item $I=1,S=0.$
	\begin{align}
		&\left(T_{bb}^{1,0}\right)^{0}=bb\bar{d}\bar{d},\label{Tbb5}\\
		&\left(T_{bb}^{1,0}\right)^{-}=-\frac{1}{\sqrt{2}}bb\left(\bar{u}\bar{d}+\bar{d}\bar{u}\right),\label{Tbb6}\\
		&\left(T_{bb}^{1,0}\right)^{--}=bb\bar{u}\bar{u},\label{Tbb7}
	\end{align}
\end{itemize}
here we use $\left(T_{bb}^{I,S}\right)^Q$ to denote $bb\bar{q}\bar{q}$ systems, where $I$, $S$ and $Q$ are isospin, strangeness, electric charge of the states, respectively.

And all possible configurations for the $bb\bar{q}\bar{q}$ systems with $J^P$ quantum numbers are given in Table~\ref{Configurations-of-double-bottom-Systems}. 
\begin{table}[h]
	\caption{Configurations of double-bottom systems.}\label{Configurations-of-double-bottom-Systems}
	\renewcommand
	\tabcolsep{0.93cm}
	\renewcommand{\arraystretch}{1.8}
	\begin{tabular}{ccc}
		\toprule
		
		$T_{bb}^{I,S}$&$J^{P}$&Configuration\\\hline
		$T_{bb}^{0,0}$&$1^{+}$&$|3\rangle$\\
		
		&       &$|6\rangle$\\\hline
		
		$T_{bb}^{0,2}$&$0^{+}$&$|1\rangle$\\
		
		&       &$|7\rangle$\\\cline{2-3}
		
		&$1^{+}$&$|7\rangle$\\\cline{2-3}
		
		&$2^{+}$&$|7\rangle$\\\hline
		
		$T_{bb}^{\frac{1}{2},1}$&$0^{+}$&$|1\rangle$\\
		
		&       &$|7\rangle$\\\cline{2-3}
		
		&$1^{+}$&$|3\rangle$\\
		
		&       &$|6\rangle$\\
		
		&       &$|7\rangle$\\\cline{2-3}
		
		&$2^{+}$&$|7\rangle$\\\hline
		
		$T_{bb}^{1,0}$&$0^{+}$&$|1\rangle$\\
		
		&       &$|7\rangle$\\\cline{2-3}
		
		&$1^{+}$&$|7\rangle$\\\cline{2-3}
		
		&$2^{+}$&$|7\rangle$\\
		\toprule
	\end{tabular}
\end{table}

	\subsubsection{Spatial wave function}

The spatial wave function of a few-body system can be expanded in terms of a set of Gaussian basis functions, which forms an approximate complete set in a finite coordinate space~\cite{Hiyama:2003cu}. In present work, all the quarks and antiquarks are considered to be in their S-states, accordingly, the wave function of the four-quark system in the coordinate space is expanded by a series of Gaussian functions as follow,
\begin{equation}
	\Psi(\{\vec{r}_i\})=\prod_{i=1}^{4}\sum_{\mathcal{l}}^{n}C_{i\mathcal{l}}\left(\frac{1}{\pi b_{i\mathcal{l}}^2}\right)^{3/4}\mathrm{exp}\left[-\frac{1}{2b_{i\mathcal{l}}^2}r_i^2\right]\,,
\end{equation}
where $\{b_{i\mathcal{l}}\}$ are the harmonic oscillator length parameters, which can be related to the angular frequencies $\{\omega_{\mathcal{l}}\}$ of the harmonic oscillator by $1/b_{i\mathcal{l}}^2=m_i\omega_{\mathcal{l}}$. Here we assume that $\{\omega_{\mathcal{l}}\}$ are independent of the quark mass, i.e.,$1/b_{i\mathcal{l}}^2=1/b_{\mathcal{l}}^2=m_i\omega_{\mathcal{l}}$, then the spatial wave function can be simplified as:
\begin{align}
	\Psi(\{\vec{r}_i\})=&\sum_{\mathcal{l}}^{n}C_{\mathcal{l}}\prod_{i=1}^{4}\left(\frac{m_i\omega_{\mathcal{l}}}{\pi }\right)^{3/4}\mathrm{exp}\left[-\frac{m_i\omega_{\mathcal{l}}}{2}r_i^2\right]\notag\\
	=&\sum_{\mathcal{l}}^{n}C_{\mathcal{l}}\,\psi\left(\omega_{\mathcal{l}},\{\vec{r}_i\}\right)\,,\label{SpatialF}
\end{align}
which is often adopted for the calculations of multiquark systems~\cite{Zhang:2007mu,Zhang:2005jz}. Following the routine in Ref.~\cite{Hiyama:2003cu}, the parameters $\{b_{\mathcal{l}}\}$ are set to be geometric series,
\begin{equation}
	b_{\mathcal{l}}=b_1a^{\mathcal{l}-1}\hspace{0.8cm}\left(\mathcal{l}=1,2,...,n\right)\,,
\end{equation}
where three parameters $\{b_{1},b_{n},n\}$ are the Gaussian size parameters in geometric progression for numerical calculations,
and the final results are stable and independent with these parameters within an approximate complete set in a sufficiently
large space.

In addition, in order to remove the influence of the center-of-mass kinetic energy, here we rewrite the tetraquark wave function using the Jacobian coordinates. For Four-body system state, the Jacobi coordinates are  defined as
\begin{equation}\vec{\xi_{1}}=\vec{r_{1}}-\vec{r_{2}},~~~\vec{\xi_{2}}=\vec{r_{3}}-\vec{r_{4}},\notag\end{equation}
\begin{equation}\vec{\xi_{3}}=\frac{m_1\vec{r_{1}}+m_2\vec{r_{2}}}{m_1+m_2}-\frac{m_3\vec{r_{3}}+m_3\vec{r_{3}}}{m_3+m_4},\notag\end{equation}
and
\begin{equation}\vec{\xi_{4}}=\frac{m_1\vec{r_{1}}+m_2\vec{r_{2}}+m_3\vec{r_{3}}+m_3\vec{r_{3}}}{m_1+m_2+m_3+m_4}.\end{equation}
Then, other relevant coordinates of this system can be expressed in terms of $\vec{\xi_{1}}$, $\vec{\xi_{2}}$, and $\vec{\xi_{3}}$. Therefore, Eq.\eqref{SpatialF} can be written as
\begin{align}
	\Psi(\{\vec{\xi}_i\})=&\sum_{\mathcal{l}}^{n}C_{\mathcal{l}}\prod_{i=1}^{4}\left(\frac{\mu_i\omega_{\mathcal{l}}}{\pi }\right)^{3/4}\mathrm{exp}\left[-\frac{\mu_i\omega_{\mathcal{l}}\xi_i^2}{2}\right]
\end{align}
where $\mu_1=\frac{m_1m_2}{m_1+m_2}$, $\mu_2=\frac{m_3m_4}{m_3+m_4}$, $\mu_3=\frac{(m_1+m_2)(m_3+m_4)}{M}$,  $\mu_4=M=m_1+m_2+m_3+m_4$.

\subsection{Calculation Method}

Once all matrix elements have been obtained, the mass spectra can be obtained by solving the generalized matrix eigenvalue problem:
\begin{equation}
	\sum_{\mathcal{l}}^n\sum_{\mathcal{l}'}^nC^i_{\mathcal{l}}\left(H^d_{\mathcal{l}\mathcal{l}'}-E_i^d N_{\mathcal{l}\mathcal{l}'}\right)C^i_{\mathcal{l}'}=0\,,\label{GEP}
\end{equation}
where $i=1\text{--}n$ and
\begin{align}
	H^d_{\mathcal{l}\mathcal{l}'}=&\langle\psi\left(\omega_{\mathcal{l}}\right)(SCF)|H_{eff.}|\psi\left(\omega_{\mathcal{l}'}\right)(SCF)\rangle\,,\\
	N_{\mathcal{l}\mathcal{l}'}=&\langle\psi\left(\omega_{\mathcal{l}}\right)(SCF)|\psi\left(\omega_{\mathcal{l}'}\right)(SCF)\rangle\,.
\end{align}
Where, the $H^d_{\mathcal{l}\mathcal{l}'}$ are the matrix elements in the total  spin-color-flavor-spatial bases, $E_i^d$ stands for the eigenvalue, $C^i_{\mathcal{l}'}$ are the relevant eigenvector and $(SCF)$ stand for spin$\,\times\,$color$\,\times\,$flavor wave function. Note that since the Gaussian function of different harmonic oscillator frequencies is not orthogonal, $N_{\mathcal{l}\mathcal{l}'}$ is not an identity matrix.
	
From TABLE~\ref{Configurations-of-hidden-bottom-Systems} and TABLE~\ref{Configurations-of-double-bottom-Systems}, a given system may include several different configurations with same $IJ^{P(C)}$, which can mix with each other. In the current computation, we first derive $n$ eigenenergy $E_i^d$ and $n$ corresponding eigenvector $C^i$ by solving Eq.\eqref{GEP} to get the masses of pure configurations, and then calculate the off-diagonal effects between different configurations.

Additionally, according to the Rayleigh-Ritz variational principle, the lowest eigenenergy $E_m^d$ should correspond to the steady-state energy. Therefore, we need to select a set of $\{b_{1},b_{n},n\}$, and then encrypt and widen the value of the harmonic oscillator length parameter until $E^d_m$ tends to be stable.  In present work, we take the parameter $\{b_{1},b_{n},n\}=\{0.02fm,6fm,40\}$ for the calculation.

\subsection{Model Parameters}

Up to now, there is no solid confirmation on existence of the present studied $q\bar{q}b\bar{b}$ and $T_{bb}$ tetraquark states, therefore, it is impossible to determine suitable values for the parameters by fitting the masses of tetraquark states. On the other hand, in the presently employed quark model, the interaction between quarks and antiquarks in tetraquark states is treated as a simple sum of two-body confinement and two-body residual interactions, which is analogous to the description of traditional mesons in constituent quark models. Consequently, we use the INS model to fit the masses of a set of traditional ground-state mesons listed in the PDG~\cite{ParticleDataGroup:2024cfk}, the resulted values of the present used model parameters are given in Table~\ref{Model parameters}, as we can see in the table, data for masses of the involved mesons is perfectly reproduced only except for that of the $\omega$ meson, for which the presently obtained mass is $\sim 1\%$ deviation from the experimental data. And the corresponding fitted mass spectra of the traditional mesons are shown in Table~\ref{Ground state meson spectrum}.

\begin{table}
	\caption{Model parameters.}\label{Model parameters}
	\renewcommand
	\tabcolsep{0.2cm}
	\renewcommand{\arraystretch}{1.8}
	\begin{tabular}{cc|cc}
		\hline\hline
		Parameter(Unit)& Value & Parameter(Unit) & Value\\\hline
		$m_n\,\,(\mathrm{MeV})$   &  $340$   &$m_s\,\,(\mathrm{MeV})$   &  $511$\\
		$m_b\,\,(\mathrm{MeV})$   &  $5000$  &$b\,\,(\mathrm{10^3MeV^2})$   &  $77$   \\
		$C_0\,\,(\mathrm{MeV})$   &  $-296$  &$\sigma\,\,(\mathrm{MeV})$      &  $485$ \\
		$\alpha_{nn}$             &  $0.600$ &$\alpha_{ns}$             &  $0.600$ \\
		$\alpha_{nb}$             &  $0.467$ &$\alpha_{ss}$             &  $0.562$ \\
		$\alpha_{sb}$             &  $0.445$ &$\alpha_{bb}$             &  $0.398$ \\
		$g_{nn}\,\,(\mathrm{10^{-6}MeV^{-2}})$ & $16.90$& $g_{ns}\,\,(\mathrm{10^{-6}MeV^{-2}})$ & $10.70$  \\
		$g_{nb}\,\,(\mathrm{10^{-6}MeV^{-2}})$ & $1.39$ &$g_{sb}\,\,(\mathrm{10^{-6}MeV^{-2}})$ & $1.05$  \\
		\hline\hline
	\end{tabular}
\end{table}

\begin{table*}
	\caption{Ground state meson spectrum. Experimental data are taken from PDG~\cite{ParticleDataGroup:2024cfk}.}\label{Ground state meson spectrum}
	\renewcommand
	\tabcolsep{0.4cm}
	\renewcommand{\arraystretch}{2.0}
	\begin{tabular}{cccccccccccc}
		\hline\hline
		Meson & $\pi$ & $\rho$ & $\omega$ & $\phi$ & $K$ & $K^*$& $\Upsilon$ & $B$ & $B^{*}$ & $B_{s}^0$ & $B_s^*$  \\
		
		$Expt.$ & $138$ & $775$ & $783$  &  $1019$ &  $496$  & $895$ & $9460$ &  $5279$  &  $5325$ &  $5367$  &  $5415$\\
		
		$Ours$ & $139$ & $775$ & $775$  &  $1019$ &  $499$  & $894$ &  $9461$  &  $5279$ & $5325$ &  $5367$   &  $5415$ \\
		\hline\hline
	\end{tabular}
\end{table*}

\section{Results and Discussions}
\label{Results and Discussions}

Using the model parameters in TABLE~\ref{Model parameters} to solve Eq.\eqref{GEP}, numerical results for the ground S-wave hidden- and double-bottom tetraquark states within the INS framework can be obtained. The numerical results are shown in Tables~\ref{X-states}, \ref{Zbstates}, \ref{Zbs} and~\ref{Tbb}, for $X_{q\bar{q}}$, $Z_b$, $Z_{bs}$ and $T_{bb}$ tetraquark states, respectively.

In general, if a tetraquark state may not be a stable state if it just lies above threshold of a meson-meson channel. Consequently, here we list the mass thresholds of the two mesons corresponding to the hidden bottom and double-bottom tetraquark states in the last column of the tables, and then try to figure out the possible stable tetraquark states.

In following, we will use the threshold scheme to discuss the results of hidden-bottom $X_{q\bar{q}}$, $Z_b$, $Z_{bs}$ and double-bottom $T_{bb}$ tetraquark states, respectively.

\subsection{Hidden-bottom Systems}
\subsubsection{$X_{q\bar{q}}$ states}

\begin{figure}[hbtp]
		\centering
		\includegraphics[scale=0.35]{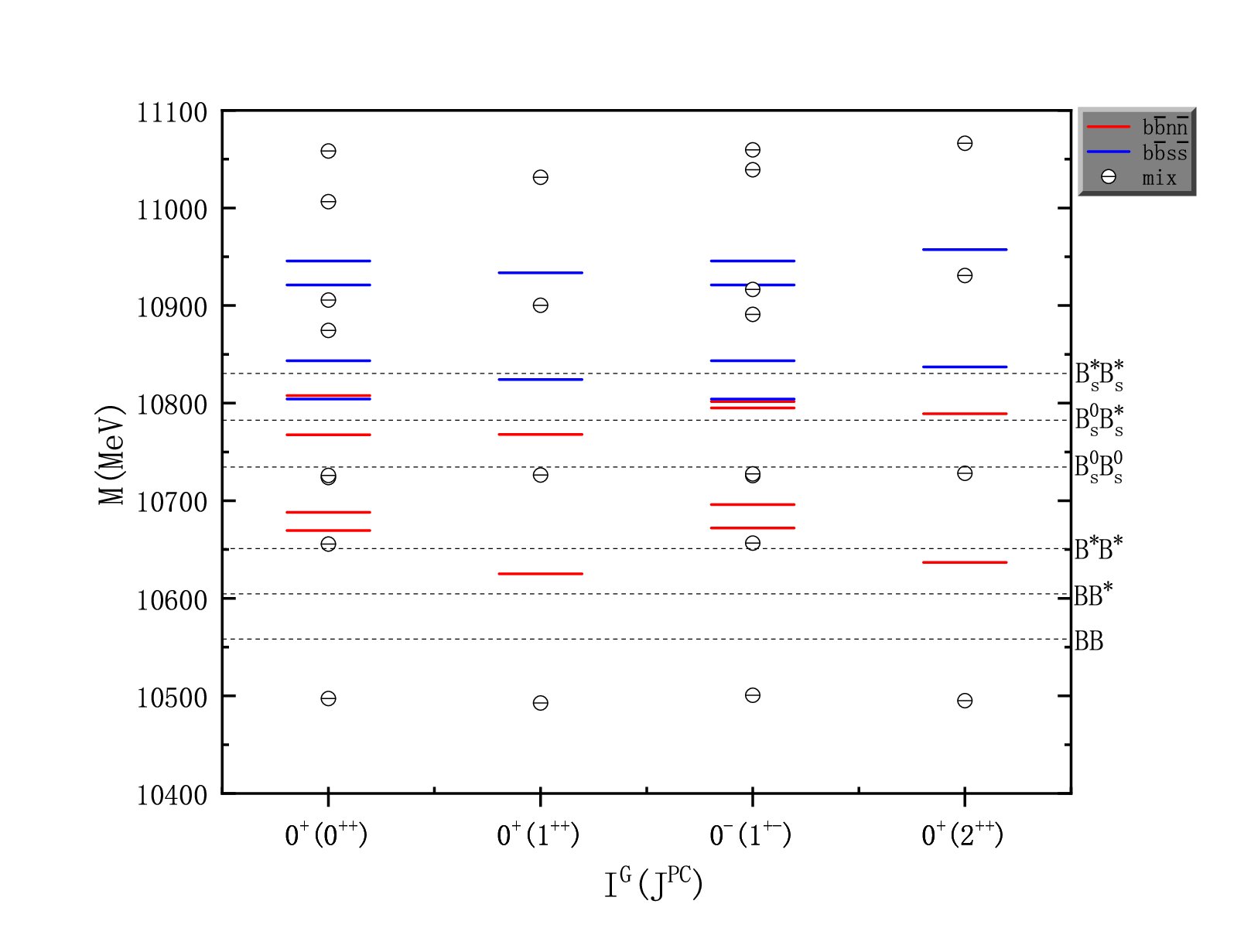}
		\caption{Spectra of obtained $X_{q\bar{q}}$ states. The red solid lines denote results for $b\bar{b}n\bar{n}$ system and blue lines denote those for the $b\bar{b}s\bar{s}$ system obtained without configuration mixing, hollow circles with a horizontal denote results after considering $b\bar{b}s\bar{s}$ and $b\bar{b}n\bar{n}$ configurations mixing, finally, the gray dotted lines are corresponding S-wave bottom meson pair thresholds.}\label{figx}
	\end{figure}

Firstly, we discuss the $b\bar{b}n\bar{n}$ and $b\bar{b}s\bar{s}$ tetraquark states, whose numerical results are shown in Table~\ref{X-states}, with the obtained energies for single configuration, the mixed configurations, respectively, the configurations mixing coefficients and the corresponding thresholds for meson-meson channels. One may notice that the instanton induced interactions should also cause mixing between the $b\bar{b}n\bar{n}$ and $b\bar{b}s\bar{s}$ configurations, therefore, we also show the numerical results in FIG.~\ref{figx}, where the red solid lines denote results for $b\bar{b}n\bar{n}$ system and blue lines denote those for the $b\bar{b}s\bar{s}$ system obtained without configuration mixing, hollow circles with a horizontal denote results after considering $b\bar{b}s\bar{s}$ and $b\bar{b}n\bar{n}$ configurations mixing, finally, the gray dotted lines are corresponding S-wave bottom meson pair thresholds.

As one can see in Table~\ref{X-states}, for the $b\bar{b}n\bar{n}$ system without configuration mixing, all the eight obtained states with quantum numbers $I^G(J^{PC})=0^+(0^{++})$ and $I^G(J^{PC})=0^-(1^{+-})$ are above the thresholds of $B^*\bar{B}^*$, $B^0\bar{B}^*$ and $B^0\bar{B}^0$, so one may conclude that there is no stable $b\bar{b}n\bar{n}$ tetraquark states without configuration mixing. Similarly, there are two obtained $b\bar{b}n\bar{n}$ states with $I^G(J^{PC})=0^+(1^{++})$, both of which are above the threshold for $B^0\bar{B}^*$. In the case of the $I^G(J^{PC})=0^+(2^{++})$ states, there are two obtained states, with one of them having the mass of $10636.8$ MeV which is below the threshold $B^*\bar{B}^*$ about $14$ MeV, while the other state lies above the threshold $B^*\bar{B}^*$ channel. 

Analogously, for the $b\bar{b}s\bar{s}$ tetraquark states, there are four states for each of $I^G(J^{PC})=0^+(0^{++})$ and $I^G(J^{PC})=0^-(1^{+-})$, and two states for each of $I^G(J^{PC})=0^+(1^{++})$ and $I^G(J^{PC})=0^+(2^{++})$. For the $0^+(0^{++})$ states, three higher states ($10843.4$ MeV, $10921.1$ MeV, $10945.6$ MeV) are above the thresholds $B^*\bar{B}^*$ about $13-115$ MeV. The lowest obtained state with mass $10804.2$ MeV, which is above the threshold $B_s^0\bar{B}_s^0$, but below the threshold $B_s^*\bar{B}_s^*$, thus it may decay to the former channel. In the case of the $0^-(1^{+-})$ states, the lowest state with mass $10815.4$ MeV, which is above the threshold $B_s^0\bar{B}_s^*$, thus may decay to $B_s^0\bar{B}_s^*$, and other states are not allowed. Finally, for states $0^+(1^{++})$ and $0^+(2^{++})$, all masses are above their corresponding thresholds, respectively.

Finally, as shown in Table~\ref{X-states}, one can find that there are several obtained $b\bar{b}n\bar{n}$ and $b\bar{b}s\bar{s}$ tetraquark states below corresponding meson-meson thresholds. For the $b\bar{b}n\bar{n}$ system, the obtained energies for stable states are $\sim10497$ MeV, $10493$ and $10501$ MeV with quantum numbers $J^{PC}=0^{++}$, $1^{++}$ and $1^{+-}$, respectively. And a stable state with $J^{PC}=2^{++}$ is also found, whose energy is $\sim10495$~MeV. For the $b\bar{b}s\bar{s}$ system, one can also find several possible stable states with $J^{PC}=0^{++}$, $1^{++}$, $1^{+-}$ and $2^{++}$, and energies for all of these states are $\sim10725$~MeV. 

In addition, as one can see in FIG.~\ref{figx}, mixing between the $b\bar{b}n\bar{n}$ and $b\bar{b}s\bar{s}$ configurations doesn't contribute too much to the final results. consequently, we suggest that one can search for the $X_{q\bar{q}}$ tetraquark states at the energies $\sim10500$~MeV and $\sim10725$~MeV.

	\begin{table*}[hbtp]
		\caption{Computed results for the $X_{q\bar{q}}$ states with $I=0$. The thresholds of two mesons are also listed.}\label{X-states}
		\renewcommand
		\tabcolsep{0.05cm}
		\renewcommand{\arraystretch}{2.0}
		\begin{tabular}{cccccccc}
			\hline\hline
			Quark content&$J^{PC}$&\multicolumn{2}{c}{\makecell{Single configuration}}& &\multicolumn{2}{c}{\makecell{Configurations mixing}}&Threshold(MeV)\\\cline{3-4}\cline{6-7}
			
			&$           $&Config.&Energies(MeV)& &Energies(MeV)&Mixing coefficients &            \\
			\hline
			$b\bar{b}n\bar{n}$&$0^{++}$&$\vert1\rangle$&$10669.5$& &$10497.3$&$(0.56,0.71,-0.34,0.25)$&$B^0\bar{B}^0=10558.2;
			B^*\bar{B}^*=10651.0$\\
			  &  &$\vert2\rangle$&$10767.6$& &$10655.5$&$(-0.25,-0.34,-0.78,0.46)$& \\
			  &  &$\vert7\rangle$&$10807.8$& &$10874.5$&$(0.38,-0.34,0.46,0.73)$&  \\
			  &  &$\vert8\rangle$&$10688.1$& &$10905.7$&$(0.70,-0.51,-0.25,-0.43)$&\\
			  &$1^{++}$&$\vert3'\rangle$&$10768.0$& &$10492.7$&$(-0.82,-0.57)$&$B^0\bar{B}^*=10604.6$\\
			  &  &$\vert5'\rangle$&$10625.0$& &$10900.3$&$(0.57,-0.82)$& \\
			  &$1^{+-}$&$\vert4'\rangle$&$10795.1$& &$10500.7$&$(0.73,0.54,0.35,-0.22)$&$B^0\bar{B}^*=10604.6;B^*\bar{B}^*=10651.0$ \\
			  &  &$\vert6'\rangle$&$10696.0$& &$10656.6$&$(0.34,0.24,-0.76,0.49)$& \\
			  &  &$\vert7\rangle$&$10801.7$& &$10890.9$&$(0.42,-0.57,-0.358,-0.60)$& \\
			  &  &$\vert8\rangle$&$10672.0$& &$10916.5$&$(0.42,-0.57,-0.358,-0.60)$& \\
			  &$2^{++}$&$\vert7\rangle$&$10789.1$& &$10495.2$&$(-0.82,0.57)$&$B^*\bar{B}^*=10651.0$ \\
			  &  &$\vert8\rangle$&$10636.8$& &$10930.7$&$(-0.57,-0.82)$& \\\hline
			  $b\bar{b}s\bar{s}$&$0^{++}$&$\vert1\rangle$&$10843.4$& &$10723.7$&$(0.56,0.71,-0.34,0.25)$&$B_s^0\bar{B}_s^0=10734.6;
			  B_s^*\bar{B}_s^*=10830.4$\\
			   &  &$\vert2\rangle$&$10921.1$& &$10725.9$&$(-0.35,-0.50,0.69,-0.38)$& \\
			  &  &$\vert7\rangle$&$10945.6$& &$11006.5$&$(0.76,-0.60,-0.15,-0.20)$& \\
			  &  &$\vert8\rangle$&$10804.2$& &$11058.3$&$(0.22,-0.12,0.49,0.83)$& \\
			  &$1^{++}$&$\vert3'\rangle$&$10933.5$& &$10726.3$&$(-0.82,-0.57)$&$B_s^0\bar{B}_s^*=10782.5$\\
			  &  &$\vert5'\rangle$&$10824.2$& &$11031.4$&$(0.57,-0.82)$& \\
			  &$1^{+-}$&$\vert4'\rangle$&$10941.6$& &$10725.8$&$(0.73,0.54,0.35,-0.22)$&$B_s^0\bar{B}_s^*=10782.5;B_s^*\bar{B}_s^*=10830.4$ \\
			  &  &$\vert6'\rangle$&$10845.5$& &$10727.5$&$(0.00,0.00,0.85,-0.53)$& \\
			  &  &$\vert7\rangle$&$10949.6$& &$11039.1$&$(0.60,0.80,0.00,0.00)$& \\
			  &  &$\vert8\rangle$&$10815.4$& &$11059.6$&$(0.00,0.00,0.53,0.85)$& \\
			  &$2^{++}$&$\vert7\rangle$&$10957.4$& &$10728.1$&$(-0.82,0.57)$&$B_s^*\bar{B}_s^*=10830.4$ \\
			  &  &$\vert8\rangle$&$10837.1$& &$11066.4$&$(-0.57,-0.82)$& \\
			  \hline\hline
				\end{tabular}
		\end{table*}
	
\subsubsection{$Z_b$ states}

\begin{figure}[hbtp]
				\centering
				\includegraphics[width=0.55\textwidth]{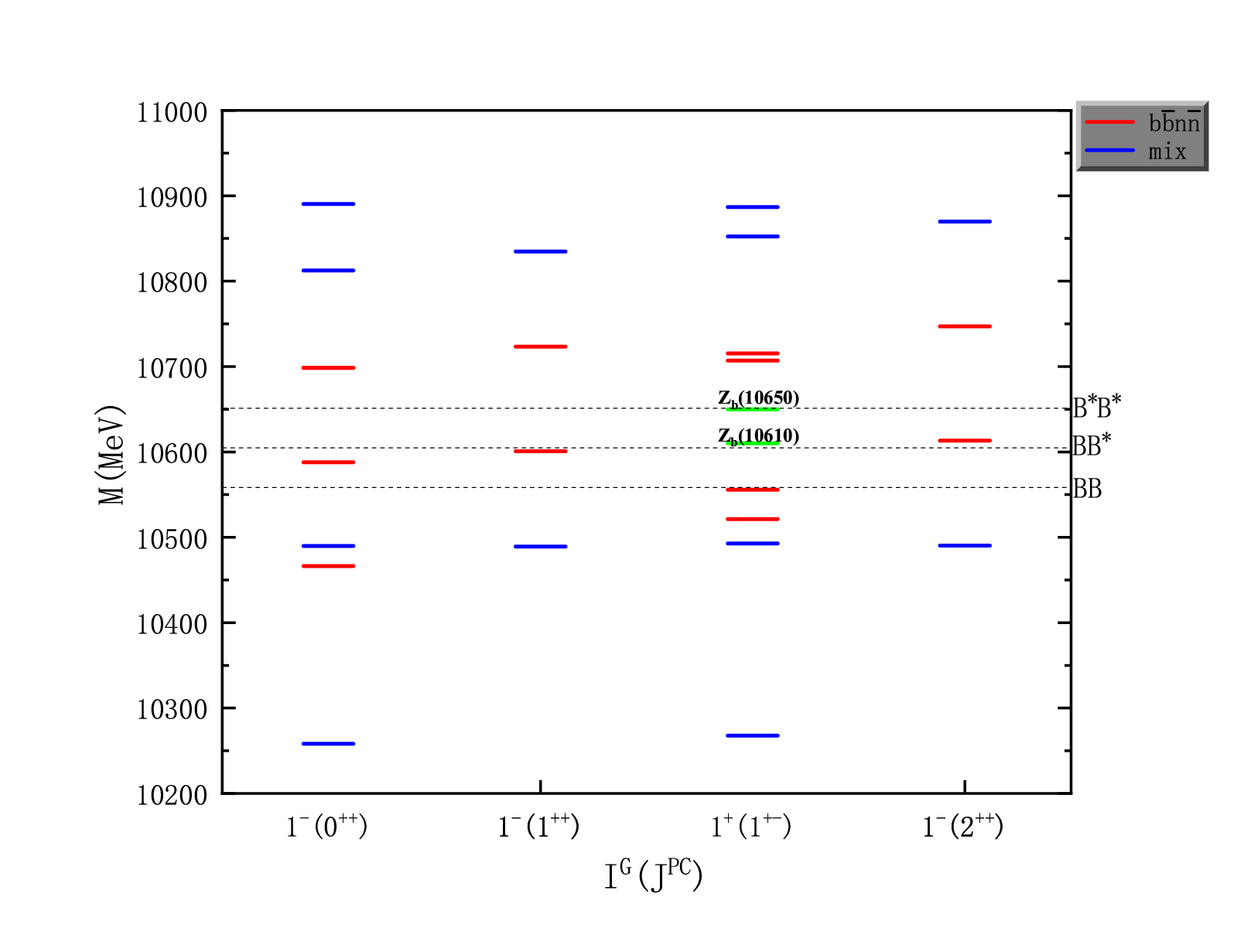}
				\caption{$Z_b$ states. The red solid lines denote energies for energies of the $b\bar{b}n\bar{n}$ states without configuration mixing, and blue lines denote results obtained by taking the effect of $b\bar{b}n\bar{n}$ configurations mixing into account, the gray dotted lines are relevant  S-wave bottom meson pair thresholds, finally, the green rectangles represent experimental masses of the observed $Z_b$ states taken from Ref.~\cite{ParticleDataGroup:2024cfk}.}\label{zbfig}
			\end{figure}

For the $b\bar{b}n\bar{n}$ tetraquarks, namely, the $Z_b$ states, we show the numerical results obtained by the single configuration, the mixed configurations, as well the configurations mixing coefficients and thresholds of relevant meson-meson channels in TABLE~\ref{Zbstates}. And we also show the numerical results in FIG.~\ref{zbfig}, where the red solid lines denote energies for energies of the $b\bar{b}n\bar{n}$ states without configuration mixing, and blue lines denote results obtained by taking the effect of $b\bar{b}n\bar{n}$ configurations mixing into account, the gray dotted lines are relevant  S-wave bottom meson pair thresholds, finally, the green rectangles represent experimental masses of the observed $Z_b$ states taken from ~\cite{ParticleDataGroup:2024cfk}.

Experimentally, in 2011, the $Z_b^\pm(10610)$ and $Z_b^\pm(10650)$ states were observed by Belle collaboration~\cite{Adachi:2011mks} and the quantum numbers for both of these two states were determined to be $I^G(J^P)=1^+(1^+)$. Explicitly, their masses and decay widths are: $M_{Z_b}(10610)=(10607.2\pm2.0)$~MeV, $M_{Z_b}(10650)=(10652.2\pm1.5)$~MeV, $\Gamma_{Z_b}(10610)= (18.4\pm2.4)$~MeV, $\Gamma_{Z_b}(10650) = (11.5\pm2.2)$~MeV~\cite{Belle:2011aa}. Obviously, their masses are extremely close to the $B^0\bar{B}^*$ and $B^*\bar{B}^*$ mass threshold, respectively. Therefore, one can expect the $B^0\bar{B}^*$ and $B^*\bar{B}^*$ hadronic molecule components should dominate these two observed states, respectively. While in present work, we also try to explore whether the compact tetraquark states can form sizable components of $Z_b^\pm(10610)$ and $Z_b^\pm(10650)$. 

As one can see in Table~\ref{Zbstates}, once the configurations mixing effects are considered, two possible stable state lies at $\sim10489$ MeV with quantum number $J^{PC}=1^{++}$, and $\sim10493$~MeV with the quantum number $1^{+-}$, respectively, are close to the $B^0\bar{B}^*$ threshold, with the deviations $\sim1\%$, namely, these two obtained compact tetraquark states may take non-negligible probabilities in $Z_b^\pm(10610)$ state. In addition, one may determine the C-parity of $Z_b^\pm(10610)$ by examinations of the decay properties of the presently obtained two compact tetraquark states.

As shown in Table~\ref{Zbstates} and FIG.~\ref{zbfig}, except for the two possible stable tetraquark states with quantum numbers $J^{PC}=1^{++(-)}$ mentioned above, there are several other states below relevant meson-meson channels' threshold. Two states with quantum numbers $J^{PC}=0^{++}$ lie at energies $\sim10258$~MeV and $10490$~MeV, respectively, especially, the latter one is below and close to the $B^{0}\bar{B}^{0}$ threshold, thus it may form a stable exotic meson state which strongly couples to $B^{0}\bar{B}^{0}$ channel. In addition, there is another state, whose energy is $\sim10490$~MeV, and quantum number $J^{PC}=2^{++}$, is also close to $B^{0}\bar{B}^{0}$ threshold. Consequently, one may search for the meson exotic states $Z_b$ around $10500$~MeV.  

		\begin{table*}[hbtp]
		\caption{Computed results for the $Z_{b}$ states with $I=1$. The thresholds of two mesons are also listed.}\label{Zbstates}
		\renewcommand
		\tabcolsep{0.1cm}
		\renewcommand{\arraystretch}{1.8}
		\begin{tabular}{cccccccc}
			\hline\hline
			Quark content&$J^{PC}$&\multicolumn{2}{c}{\makecell{Single configuration}}& &\multicolumn{2}{c}{\makecell{Configurations mixing}}&Threshold(MeV)\\\cline{3-4}\cline{6-7}
			
			&$           $&Config.&Energies(MeV)& &Energies(MeV)&Mixing coefficients &            \\
			\hline
			$b\bar{b}n\bar{n}$&$0^{++}$&$\vert1\rangle$&$10587.9$& &$10258.1$&$(0.43,0.55,-0.60,0.39)$&$B^0\bar{B}^0=10558.2;
			B^*\bar{B}^*=10651.0$\\
			&  &$\vert2\rangle$&$10698.4$& &$10489.9$&$(0.42,0.59,0.60,-0.35)$& \\
			&  &$\vert7\rangle$&$10698.6$& &$10812.6$&$(0.71,-0.52,-0.24,-0.42)$&  \\
			&  &$\vert8\rangle$&$10466.2$& &$10890.5$&$(0.37,-0.30,0.47,0.74)$&\\
			&$1^{++}$&$\vert3'\rangle$&$10723.3$& &$10489.2$&$(-0.82,-0.57)$&$B^0\bar{B}^*=10604.6$\\
			&  &$\vert5'\rangle$&$10600.7$& &$10834.8$&$(0.57,-0.82)$& \\
			&$1^{+-}$&$\vert4'\rangle$&$10707.0$& &$10267.6$&$(0.74,-0.54,-0.34,0.21)$&$B^0\bar{B}^*=10604.6;B^*\bar{B}^*=10651.0$ \\
			&  &$\vert6'\rangle$&$10555.7$& &$10492.7$&$(-0.32,0.24,-0.76,0.50)$& \\
			&  &$\vert7\rangle$&$10715.4$& &$10852.4$&$(-0.43,-0.57,-0.39,-0.59)$& \\
			&  &$\vert8\rangle$&$10521.3$& &$10886.8$&$(0.41,0.57,-0.39,-0.60)$& \\
			&$2^{++}$&$\vert7\rangle$&$10747.0$& &$10490.3$&$(-0.82,0.57)$&$B^*\bar{B}^*=10651.0$ \\
			&  &$\vert8\rangle$&$10613.2$& &$10869.8$&$(-0.57,-0.82)$& \\
			\hline\hline
		\end{tabular}
	
	\end{table*}

\subsubsection{$Z_{bs}$ states}

\begin{figure}[hbtp]
	\centering
	\includegraphics[width=0.55\textwidth]{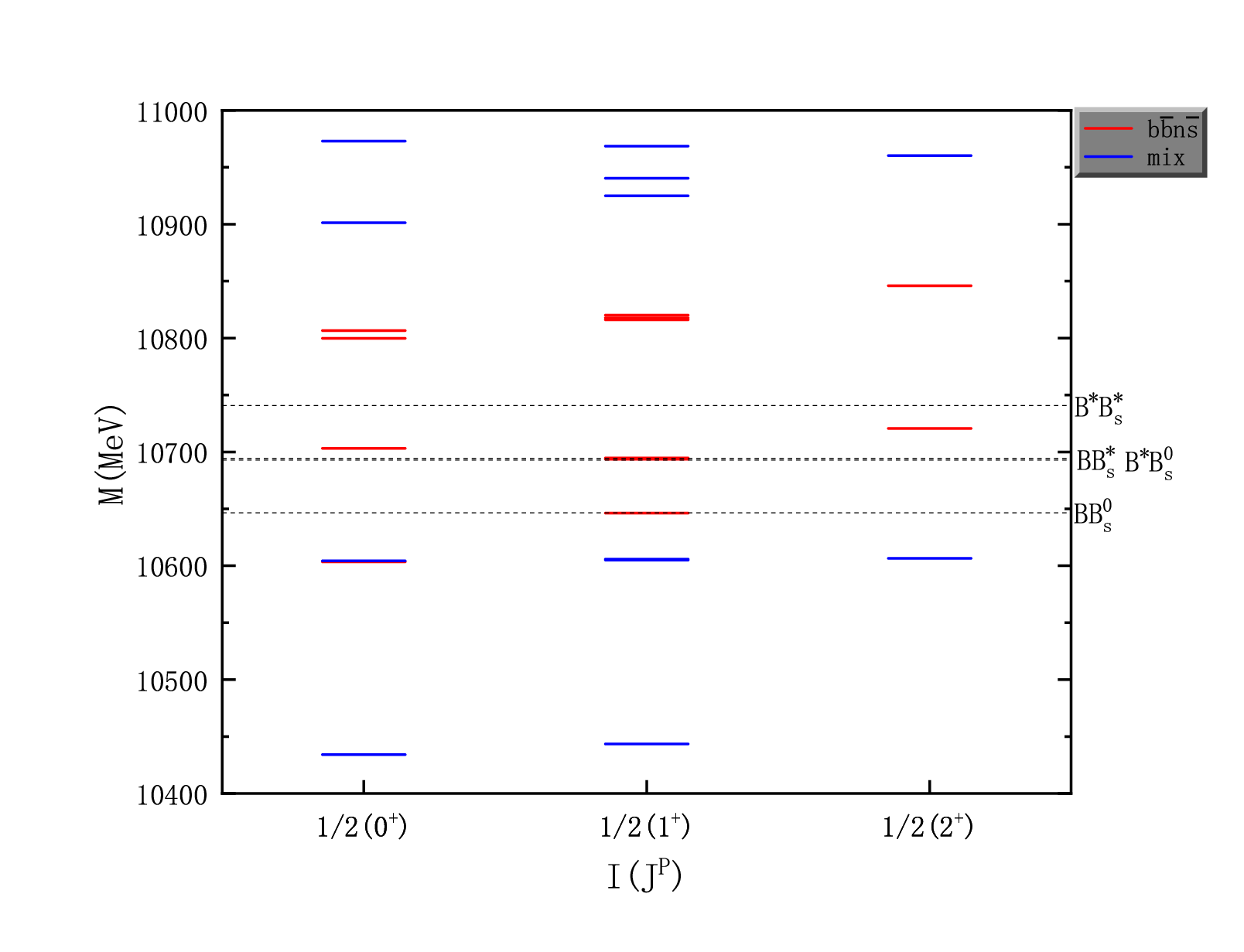}
	\caption{$Z_{bs}$ states. The red solid lines denote energy for each single $b\bar{b}n\bar{s}$ configuration, while the blue lines denote the numerical results after taking the effects of configuration mixing into account, and the gray dotted lines are relevant S-wave bottom meson pair thresholds.}\label{zbsfig}
\end{figure}

Finally, for the $b\bar{b}n\bar{s}$ tetraquarks, namely, the $Z_{bs}$ states, we show the numerical results obtained by the single configuration, and mixed configurations, as well the configurations mixing coefficients and thresholds of relevant meson-meson channels in TABLE~\ref{Zbs}. In addition, we also show the numerical results in FIG.~\ref{zbsfig}, where the red solid lines denote energy for each single $b\bar{b}n\bar{s}$ configuration, while the blue lines denote the numerical results after taking the effects of configuration mixing into account, and the gray dotted lines are relevant S-wave bottom meson pair thresholds.
	
As we can see in Table~\ref{Zbs} and FIG.~\ref{zbsfig}, once the configuration mixing effects are considered, there are two deeply bounded states whose energies are $\sim10440$~MeV, and quantum numbers are $0^+$ and $1^+$, if they are related to the $B\bar{B}_{s}$ channel, the binding energy is $\sim200$~MeV. There are four states lie at energy $\sim10605$~MeV, with quantum numbers $J^{P}=0^+$, $1^{+}$, $1^{+}$ and $2^{+}$, respectively, they are close to the $B\bar{B}_{s}$ threshold, so one may expect them to form stable exotic meson states.
	
	\begin{table*}[hbtp]
		\caption{Computed results for the $Z_{bs}$ states with $I=\frac{1}{2}$. The thresholds of two mesons are also listed.}\label{Zbs}
		\renewcommand
		\tabcolsep{0.02cm}
		\renewcommand{\arraystretch}{1.8}
		\begin{tabular}{cccccccc}
			\hline\hline
			Quark content&$J^{PC}$&\multicolumn{2}{c}{\makecell{Single configuration}}& &\multicolumn{2}{c}{\makecell{Configurations mixing}}&Threshold(MeV)\\\cline{3-4}\cline{6-7}
			
			&$           $&Config.&Energies(MeV)& &Energies(MeV)&Mixing coefficients &            \\
			\hline
			$b\bar{b}n\bar{s}$&$0^{+}$&$\vert1\rangle$&$10703.1$& &$10434.2$&$(0.45,-0.56,-0.58,0.38)$&$B^0\bar{B}_s^0=10646.4;B^*\bar{B}_s^*=10740.7$ \\
			&  &$\vert2\rangle$&$10799.8$& &$10604.3$&$(0.40,-0.57,0.62,-0.36)$&\\
			&  &$\vert7\rangle$&$10806.5$& &$10901.4$&$(-0.71,-0.52,0.23,0.42)$&\\
			&  &$\vert8\rangle$&$10603.5$& &$10973.0$&$(-0.37,-0.31,-0.47,-0.74)$&\\
			&$1^{+}$&$\vert3\rangle$&$10816.0$& &$10443.5$&$(0.52,-0.54,-0.38,0.40,-0.30,0.19)$&$B^*\bar{B}_s^0=10692.8;B^*\bar{B}_s^*=10740.7$ \\
			&  &$\vert4\rangle$&$10817.7$& &$10605.0$&$(-0.25,0.16,0.18,-0.12,-0.78,0.51)$&$B^0\bar{B}_s^*=10694.3$ \\
			&  &$\vert5\rangle$&$10694.6$& &$10605.9$&$(-0.58,-0.59,0.40,0.40,0.07,-0.04)$& \\
			&  &$\vert6\rangle$&$10693.7$& &$10924.9$&$(0.31,-0.30,0.41,-0.40,0.38,0.59)$& \\
			&  &$\vert7\rangle$&$10820.1$& &$10940.4$&$(0.40,0.40,0.58,0.59,0.00,0.01)$& \\
			&  &$\vert8\rangle$&$10646.2$& &$10968.6$&$(-0.29,0.29,-0.40,0.40,0.39,0.60)$& \\
			&$2^{+}$&$\vert7\rangle$&$10846.0$& &$10606.5$&$(-0.82,-0.57)$&$B^*\bar{B}_s^*=10740.7$ \\
			&  &$\vert8\rangle$&$10720.6$& &$10960.2$&$(0.57,-0.82)$& \\
			\hline\hline
		\end{tabular}
	\end{table*}\textsf{}

	\subsection{Double-bottom Systems}
	
	\begin{table*}[hbtp]
		\caption{Computed results for the $T_{bb}$ states.}\label{Tbb}
		\renewcommand
		\tabcolsep{0.4cm}
		\renewcommand{\arraystretch}{1.8}
		\begin{tabular}{cccccccc}
			\hline\hline
			States&Quark content&$J^{P}$&\multicolumn{2}{c}{\makecell{Single configuration}}& &\multicolumn{2}{c}{\makecell{Configurations mixing}}\\\cline{4-5}\cline{7-8}
			
			&  &  &Config.&Energies(MeV)& &Energies(MeV)&Mixing coefficients\\
			\hline
			$T_{bb}^{0,0}$&$bb\bar{n}\bar{n}$&$1^{+}$&$\vert3\rangle$&$10558.0$& &$10557.0$&$(-0.06,1.00)$\\
			&  &  &$\vert6\rangle$&$10881.8$& &$10882.8$&$(-1.00,-0.06)$\\\hline
			$T_{bb}^{1,0}$&$bb\bar{n}\bar{n}$&$0^{+}$&$\vert1\rangle$&$10924.2$& &$10647.0$&$(-1.00,-0.10)$\\
			&  &  &$\vert7\rangle$&$10649.7$& &$10926.9$&$(0.10,-1.00)$\\
			&  &$1^{+}$&$\vert7\rangle$&$10657.8$& &$-$&$-$\\
			&  &$2^{+}$&$\vert7\rangle$&$10673.5$& &$-$&$-$\\\hline
			$T_{bb}^{{\frac{1}{2}},1}$&$bb\bar{n}\bar{s}$&$0^{+}$&$\vert1\rangle$&$11002.6$& &$10746.8$&$(-0.99,0.11)$\\
			&  &  &$\vert7\rangle$&$10750.0$& &$11005.7$&$(-0.11,-0.99)$\\
			&  &$1^{+}$&$\vert3\rangle$&$10686.5$& &$10685.4$&$(-0.06,-1.00,0.01)$\\
			&  &  &$\vert6\rangle$&$10970.3$& &$10758.3$&$(0.01,0.01,1.00)$\\
			&  &  &$\vert7\rangle$&$10758.3$& &$10971.4$&$(1.00,-0.06,-0.01)$\\
			&  &$2^{+}$&$\vert7\rangle$&$10774.7$& &$-$&$-$\\\hline
			$T_{bb}^{0,2}$&$bb\bar{s}\bar{s}$&$0^{+}$&$\vert1\rangle$&$11079.7$& &$10848.6$&$(-0.99,-0.13)$\\
			&  &  &$\vert7\rangle$&$10852.3$& &$11083.4$&$(0.13,-0.99)$\\
			&  &$1^{+}$&$\vert7\rangle$&$10860.9$& &$-$&$-$\\
			&  &$2^{+}$&$\vert7\rangle$&$10877.7$& &$-$&$-$\\
			\hline\hline
		\end{tabular}
	\end{table*}
	
	\begin{table*}
		\caption{ Numerical results for the $T_{bb}$ states compared to those predicted by other approaches. The Columns denoted by QM are predictions by various of different quark model, and the last column is the predictions by LQCD.}\label{com}
		\renewcommand
		\tabcolsep{0.1cm}
		\renewcommand{\arraystretch}{1.8}
		\begin{tabular}{ccccccccccccc}
			\hline\hline
			States&Quark content&$J^{P}$&Ours&QM\cite{Cheng:2020wxa}&QM\cite{Luo:2017eub}&QM.\cite{Ebert:2007rn}&QM\cite{Deng:2018kly}&QM\cite{Lu:2020rog}&QM\cite{Weng:2021hje}&QM\cite{Song:2023izj}&QM\cite{Zhang:2021yul}&LQCD\cite{Braaten:2020nwp}\\
			$T_{bb}^{0,0}$&$bb\bar{n}\bar{n}$&$1^{+}$&$10558.0$&$10487.9$&$10686$&$10502$&$10282$&$10550$&$10291.6/10390.9$&$10530$&$10654$&$10471$\\
			&  &  &$10881.8$&$10621.0$&$10821$&$-$&$-$&$10951$&$10703.4/10950.3$&$-$&$10982$&$-$\\\hline
			$T_{bb}^{1,0}$&$bb\bar{n}\bar{n}$&$0^{+}$&$10924.2$&$10641.7$&$10937$&$10648$&$10558$&$11019$&$10468.8/10569.3$&$10726$&$11092$&$10664$\\
			&  &  &$10649.7$&$10737.6$&$10841$&$-$&$-$&$10765$&$10808.9/11054.6$&$-$&$10834$&$-$\\
			&  &$1^{+}$&$10657.8$&$10676.3$&$10875$&$10657$&$10586$&$10779$&$10485.3/10584.2$&$10733$&$10854$&$10671$\\
			&  &$2^{+}$&$10673.5$&$10698.7$&$10897$&$10673$&$10572$&$10799$&$10507.9/10606.8$&$10747$&$10878$&$10685$\\\hline
			$T_{bb}^{{\frac{1}{2}},1}$&$bb\bar{n}\bar{s}$&$0^{+}$&$11002.6$&$10755.4$&$11095$&$10802$&$10716$&$11098$&$10586.4/10684.1$&$10855$&$11160$&$10781$\\
			&  &  &$10750.0$&$10851.2$&$10999$&$-$&$-$&$10883$&$10854.6/11090.2$&$-$&$10955$&$-$\\
			&  &$1^{+}$&$10686.5$&$10670.9$&$10911$&$10809$&$10629$&$10734$&$10473.1/10569.0$&$10861$&$10974$&$10788$\\
			&  &  &$10970.3$&$10767.2$&$11037$&$-$&$-$&$11046$&$10605.3/10700.5$&$-$&$11068$&$-$\\
			&  &  &$10758.3$&$10794.1$&$11010$&$-$&$-$&$10897$&$10778.7/11016.1$&$-$&$10811$&$-$\\
			&  &$2^{+}$&$10774.7$&$10817.6$&$11060$&$10823$&$10734$&$10915$&$10628.7/10723.9$&$10873$&$10997$&$10802$\\\hline
			$T_{bb}^{0,2}$&$bb\bar{s}\bar{s}$&$0^{+}$&$11079.7$&$10857.7$&$11254$&$10932$&$10866$&$11155$&$10697.1/10792.1$&$10976$&$11232$&$10898$\\
			&  &  &$10852.3$&$10954.3$&$11157$&$-$&$-$&$10972$&$10928.8/11154.3$&$-$&$11078$&$-$\\
			&  &$1^{+}$&$10860.9$&$10901.0$&$11199$&$10939$&$10875$&$10986$&$10718.2/10809.8$&$10981$&$11099$&$10898$\\
			&  &$2^{+}$&$10877.7$&$10925.6$&$11224$&$10950$&$10882$&$11004$&$10742.5/10834.1$&$10991$&$11119$&$10919$\\
			\hline\hline
		\end{tabular}
	\end{table*}

	\begin{table*}
		\caption{Stability of the double-bottom tetraquarks in various studies. The meanings of "S", "US", are "stable", "unstable", respectively.}\label{sta}
		\renewcommand
		\tabcolsep{0.8cm}
		\renewcommand{\arraystretch}{1.8}
		\begin{tabular}{cccc}
			\hline\hline
			Reference&$bb\bar{n}\bar{n}$&$bb\bar{n}\bar{s}$&$bb\bar{s}\bar{s}$\\\hline
			This work&$S$&$S$&$US$\\
			Ref.\cite{Cheng:2020wxa}&$S$&$S$&$US$\\
			Ref.\cite{Luo:2017eub}&$S$&$S$&$US$\\
			Ref.\cite{Ebert:2007rn}&$S$&$US$&$US$\\
			Ref.\cite{Eichten:2017ffp}&$S$&$S$&$US$\\
			Ref.\cite{Deng:2018kly}&$S$&$S$&$US$\\
			Ref.\cite{Lu:2020rog}&$S$&$US$&$US$\\
			Ref.\cite{Du:2012wp}&$S$&$S$&$S$\\
			Ref.\cite{Braaten:2020nwp}&$S$&$S$&$US$\\
			Ref.\cite{Lee:2009rt}&$S$&$S$&$-$\\
			\hline\hline
		\end{tabular}
	\end{table*}
	
		\begin{figure}[hbtp]
		\centering
		\includegraphics[width=0.55\textwidth]{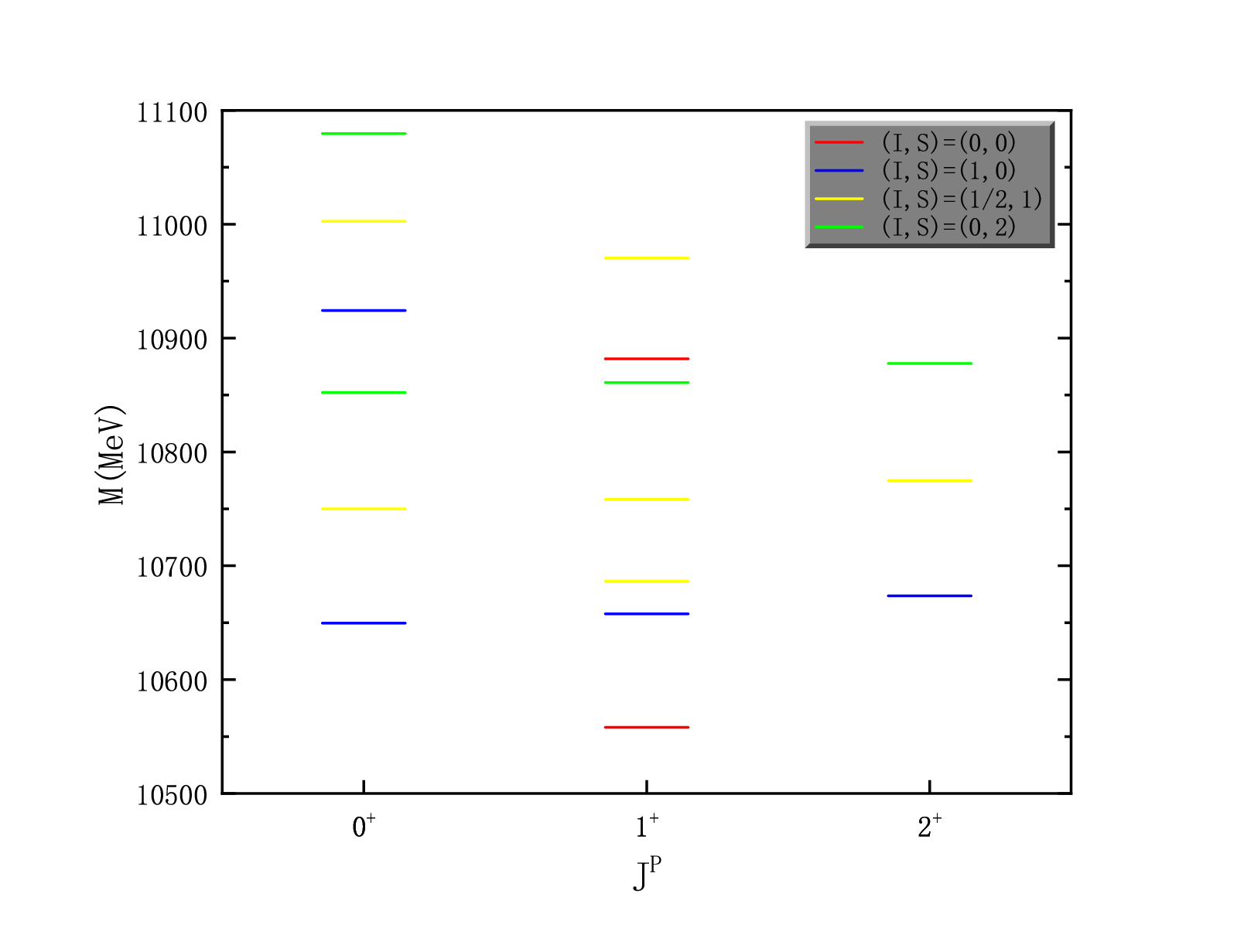}
		\caption{$T_{bb}$ states. The red solid lines denote $T^{0,0}_{bb}$ states, the blue solid lines denote $T^{1,0}_{bb}$ states, the yellow solid lines denote $T^{\frac{1}{2},1}_{bb}$ states, and the green solid line denote $T^{0,2}_{bb}$ states.}\label{Tbbfig}
	\end{figure}
	
	\begin{figure}[hbtp]
		\centering
		\includegraphics[width=0.55\textwidth]{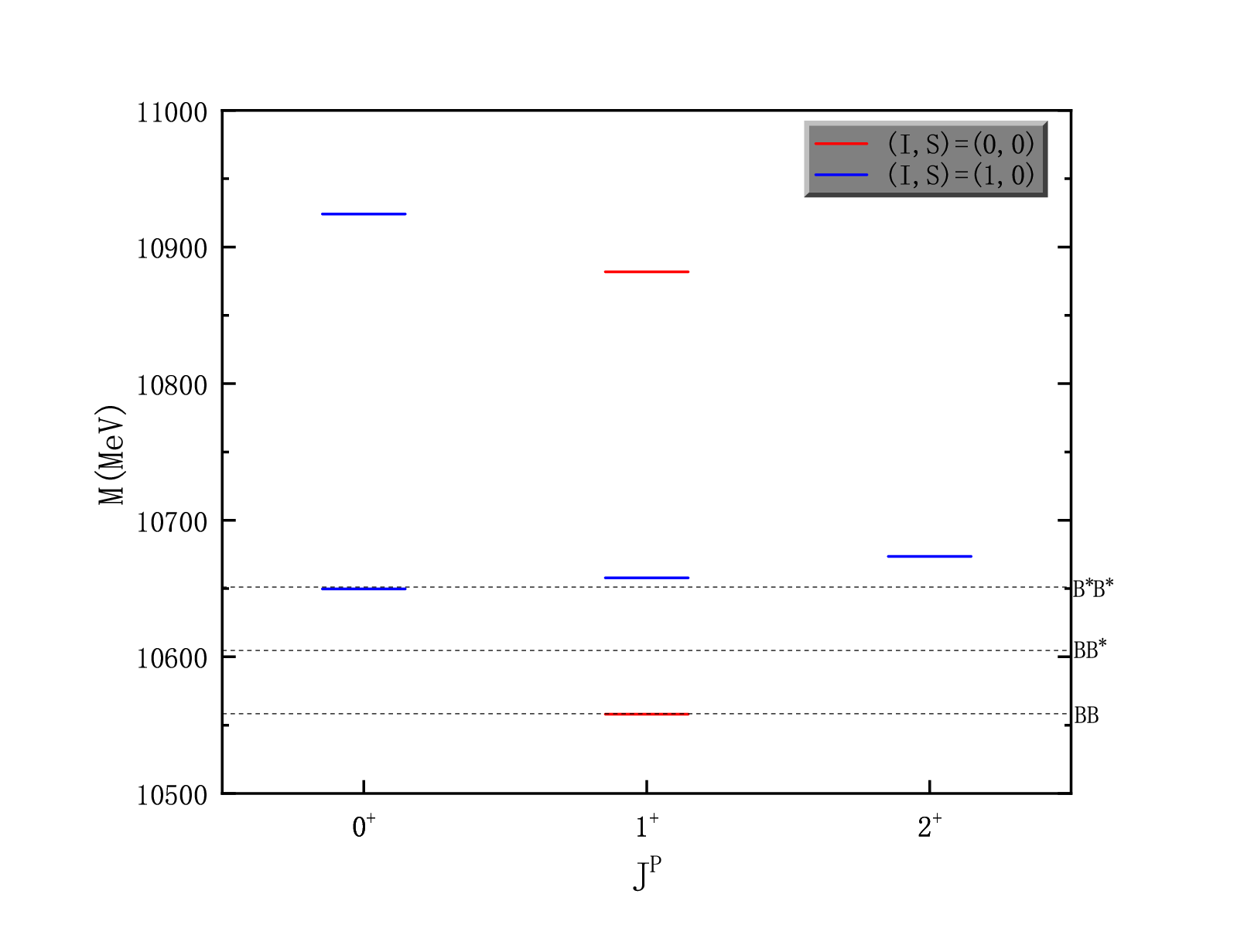}
		\caption{$bb\bar{n}\bar{n}$ states. The red solid lines denote the states with isospin $I=1$, while the blue ones are results for $I=0$, and the grey dotted lines are the thresholds of relevant meson-meson channels.}\label{bbnn}
	\end{figure}
	
	\begin{figure}[hbtp]
		\centering
		\includegraphics[width=0.55\textwidth]{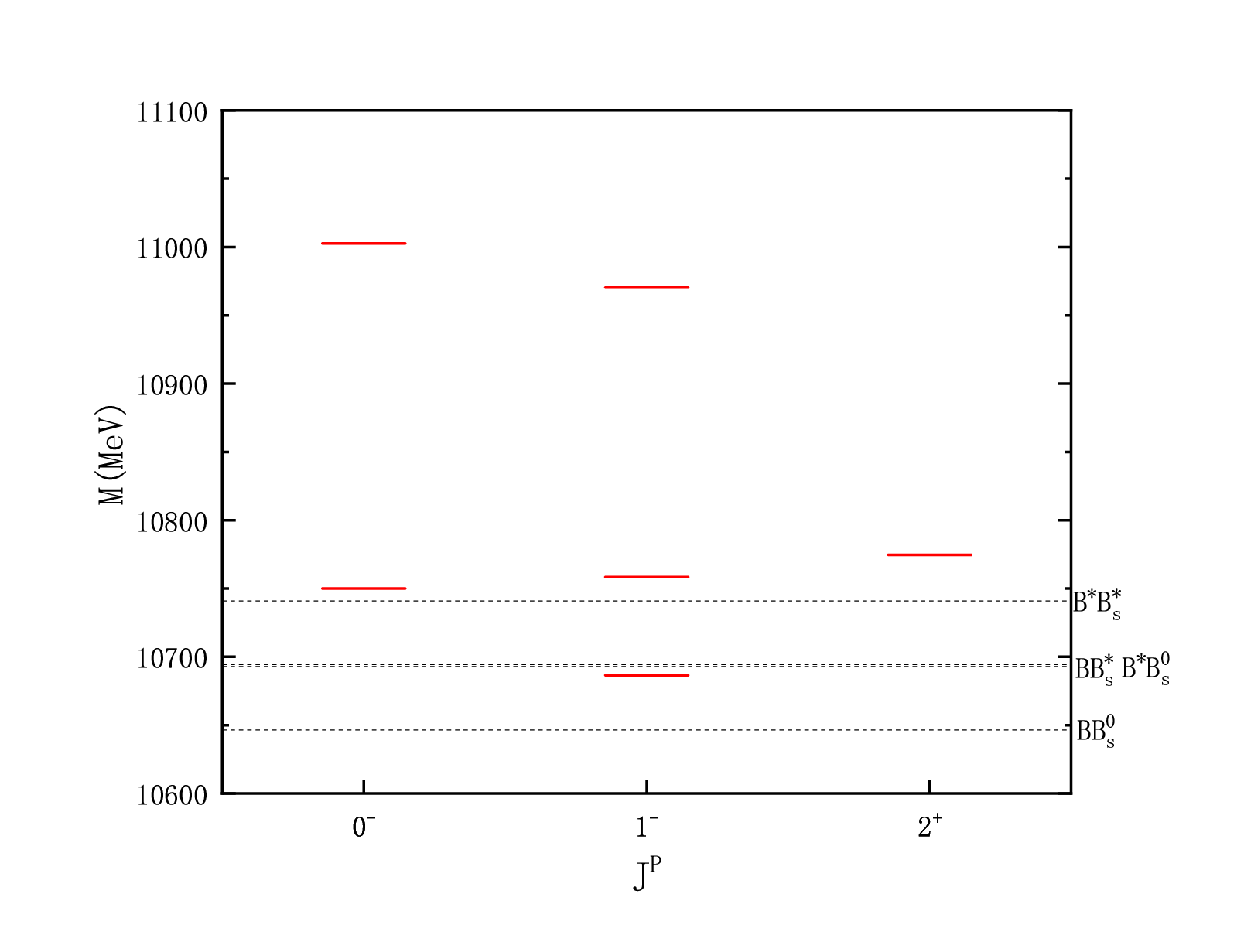}
		\caption{$bb\bar{n}\bar{s}$ states.}\label{bbns}
	\end{figure}
	
	\begin{figure}[hbtp]
		\centering
		\includegraphics[width=0.55\textwidth]{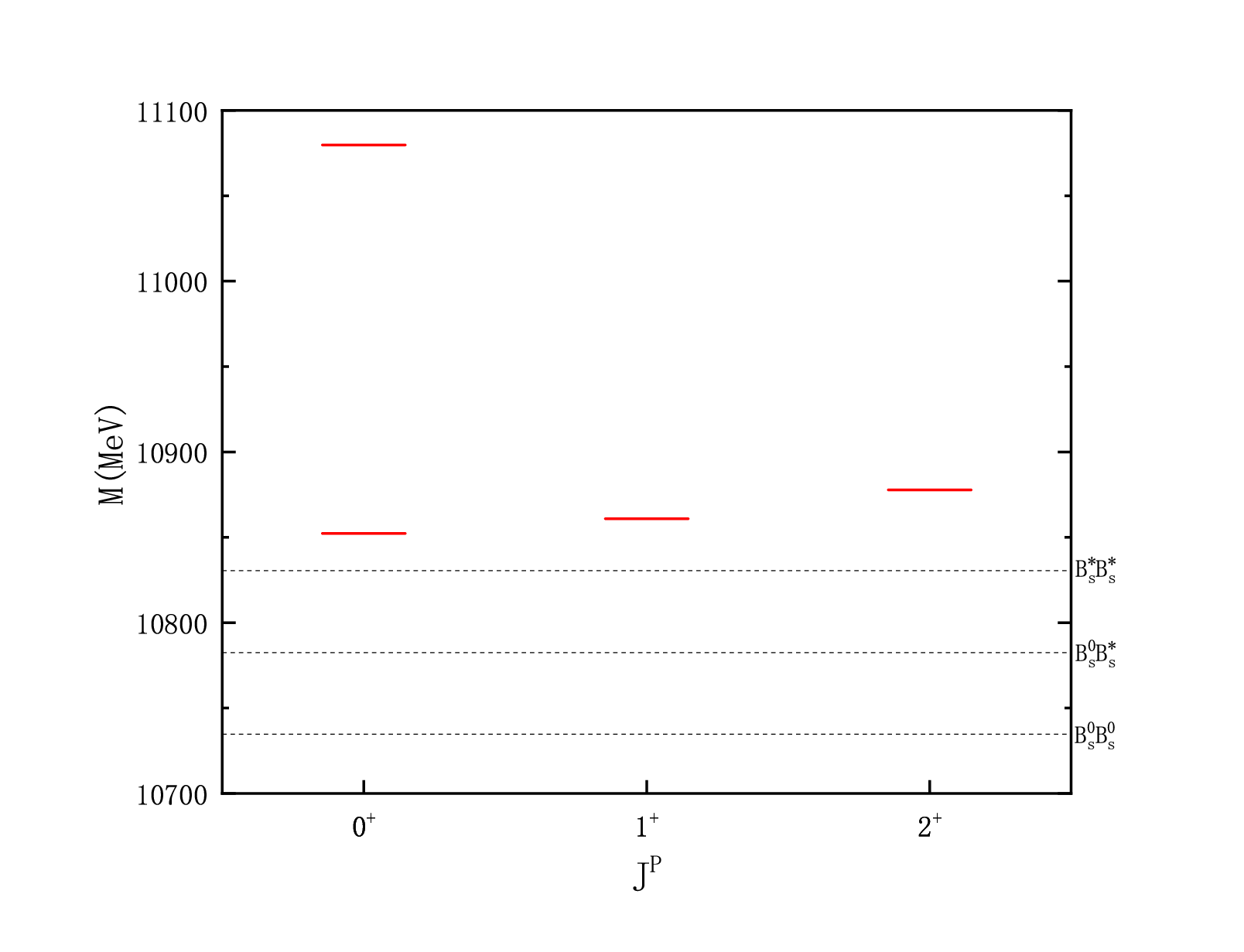}
		\caption{$bb\bar{s}\bar{s}$ states.}\label{bbss}
	\end{figure}
	
Here we come to the $bb\bar{q}\bar{q}$ tetraquarks, namely, the $T_{bb}$ states. The numerical results for energies of $bb\bar{q}\bar{q}$ states with quantum numbers $J^P=0^+$, $1^+$, and $2^+$ are shown in Table~\ref{Tbb}, where the first three columns are the labels of these states in the model, quark content and quantum number of each state respectively, the fourth, fifth columns show the results of single configuration calculations, and the last two columns are the results obtained by considering configurations mixing effects. As shown in Table~\ref{Tbb}, the presently obtained energies of the $T_{bb}$ states are in the range $10550-11100$~MeV.

The spectrum of S-wave double-bottom tetraquark states are also shown in FIG.~\ref{Tbbfig}, where the red solid lines denote $T^{0,0}_{bb}$ states, the blue solid lines denote $T^{1,0}_{bb}$ states, the yellow solid lines denote $T^{\frac{1}{2},1}_{bb}$ states, and the green solid line denote $T^{0,2}_{bb}$ states. In addition, we also show the numerical results for explicit $bb\bar{q}\bar{q}$ states with $\bar{q}\bar{q}$ being $\bar{n}\bar{n}$ ($n=u,d$), $\bar{n}\bar{s}$, and $\bar{s}\bar{s}$ in FIG.~\ref{bbnn}, FIG.~\ref{bbns} and FIG.~\ref{bbss}, respectively. In FIG.~\ref{bbnn}, the red solid lines denote the states with isospin $I=1$, while the blue ones are results for $I=0$, and the grey dotted lines are the thresholds of relevant meson-meson channels. In FIGs.~\ref{bbns}~and~\ref{bbss}, we show the presently obtained results by the red solid lines, and thresholds of relevant meson-meson channels by grey dotted lines. And in all these four figures, we just show the numerical results obtained by taking the configuration mixing effects into account.

As shown in FIG.~\ref{bbnn}, two possible stable tetraquark states are found, with the quantum numbers $J^{P}=0^{+}$, and $1^{+}$, respectively. The $T_{bb}$ state with quantum number $0^{+}$ lies at $\sim10650$~MeV, which is below and very close to the threshold of $B^*B^*$ channel. While energy of the other $T_{bb}$ state with $J^{P}=1^{+}$ is $\sim10558$~MeV, below and very close to $BB$ channel. In FIG.~\ref{bbns}, it's shown that only one possible stable tetraquark state could be found, whose energy is $\sim10687$~MeV, below and close to threshold of $BB_s^*$ channel. Finally, there is no possible stable $T_{bb}$ tetraquark state with the quark content $bb\bar{s}\bar{s}$, as shown in FIG.~\ref{bbss}.

In addition to the above discussions, we also compare the presently obtained results with predictions by other theoretical approaches in Table~\ref{com}, since there is no solid experimental data about the $T_{bb}$ states. As shown in this table, predictions by various of quark model and lattice QCD simulation are in a relatively large range, with a deviation $<3\%$ from the presently obtained numerical results.  
And in Table~\ref{sta}, the predictions on whether there is stable $T_{bb}$ tetraquark states by different theoretical approaches are shown, as one can see, the present results are in general consistent with the other theoretical predictions.	

\section{Summary}
\label{Summary}

We have performed a systematic study of the S-wave hidden-bottom and double-bottom tetraquark states in the constituent quark model, where the instanton-induced interaction is taken as the residual spin-dependent hyperfine interaction between quarks. The parameters in the INS model are obtained by fitting the known meson spectra, and are used directly to predict the mass spectra of the hidden-bottom and double-bottom tetraquark states.

For the $X_{q\bar{q}}$ states with quark content being $b\bar{b}q\bar{q}$, we suggest that one can try to search this kind of meson exotic states at the energies $\sim10500$~MeV and $10725$~MeV. In the case of $Z_b$ states, two tetraquark states with quantum number $J^{P}=1^{+}$ are found to be possible components of the experimentally observed $Z_{b}(10610)$ states, in addition, a very possible stable compact tetraquark state with quantum number $J^{PC}=2^{++}$ is found, whose energy is $\sim10490$~MeV, which is below and very close to the threshold of $B^{0}\bar{B^{0}}$ channel. And we also find four possible stable $Z_{bs}$ states, which lie at $10605$~MeV, which is below and close to the $B\bar{B}_s$ threshold.

And for the $S-$wave $bb\bar{q}\bar{q}$ tetraquark states, namely, the $T_{bb}$ states, three likely stable states are found in present work. Two of them are with the quark content $bb\bar{n}\bar{n}$~($n=u,d$), whose quantum number are $0^{+}$~and~$1^{+}$, respectively, and mass of them are $\sim10650$~MeV and $10558$~MeV, below and very close to the $B^*B^*$ and $BB$ thresholds, respectively. The other possible stable state is with the quark content $bb\bar{n}\bar{s}$, whose quantum number is $J^{P}=1^{+}$, and energy is $\sim10687$~MeV, that is below and close to the $BB^{*}_s$ threshold. Finally, there is no possible stable $T_{bb}$ state with quark content $bb\bar{s}\bar{s}$ found in present work.

\section{acknowledgements}\label{sec:acknowledgements}

This work is partly supported by the Fundamental Research Funds for the Central Universities under Grant No. SWU-XDJH202304, by the National Key R\&D Program of China under Grant No. 2023YFA1606703, by the National Natural Science Foundation of China under Grant Nos. 12475081, 12075133, 12075288, 12435007, 12361141819, and by the Natural Science Foundation of Shandong Province under Grant Nos. ZR2022ZD26. It is also supported by Taishan Scholar Project of Shandong Province (Grant No.tsqn202103062) and the Youth Innovation Promotion Association CAS.

\end{document}